\documentclass[sigconf,nonacm]{acmart}

\usepackage{csquotes}
\usepackage{multicol}
\usepackage{listings}
\usepackage[table]{xcolor}
\usepackage{booktabs}
\usepackage{collcell}
\usepackage{tikz}
\usepackage{xstring}
\usepackage{tabularx}
\usepackage{underscore}

\definecolor{codegreen}{rgb}{0,0.6,0}
\definecolor{codegray}{rgb}{0.5,0.5,0.5}
\definecolor{codepurple}{rgb}{0.58,0,0.82}
\definecolor{backcolour}{rgb}{0.95,0.95,0.92}

\lstdefinestyle{format}{
    breakatwhitespace=false,
    breaklines=true,
    captionpos=b,
    numbersep=8pt,
    tabsize=2,
    columns=fullflexible
}

\AtBeginDocument{%
  }

\setcopyright{none}
\renewcommand\footnotetextcopyrightpermission[1]{}
\renewcommand{\shortauthors}{}

\begin{document}

\title[Whale You Go Green?]{%
Whale You Go Green? Immediate and Longitudinal Effects of Embodied
Conversational AI Avatars on Sustainable Behavioral Tendencies and
Pro-Environmental Attitudes}


\begin{abstract}
Embodied conversational AI can be persuasive, but can it motivate people to develop sustainable behavioral tendencies and increase pro-environmental attitudes? We conducted a randomized experiment ($N=576$) comparing four embodied avatars---a beluga whale, Greta Thunberg, Emma Watson, and a generic human---with a text chatbot and no-intervention control. All embodied conditions increased immediate sustainable product selection relative to the control ($\beta=.401$--$.609$, all $q\leq.0126$) and chatbot ($\beta=.398$--$.604$, all $q\leq.007$), without differing significantly from one another. Only the Beluga condition increased immediate pro-environmental attitudes relative to the control after correction. Among 287 returners assessed two months later, following an intermediate assigned-condition session, Beluga also showed significantly higher pre-booster sustainable product choice compared to the control ($\beta=.657$, $q=.0176$). Although it did not significantly outperform other embodied agents, these findings establish the immediate potential of nonhuman embodied avatar interfaces as a promising direction for memorable environmental interventions.

\end{abstract}


\author{Rachel Poonsiriwong}
\affiliation{%
  \institution{MIT Media Lab}
  \city{Cambridge}
  \state{Massachusetts}
  \country{USA}
}
\email{rachelpo@media.mit.edu}

\author{Pat Pataranutaporn}
\affiliation{%
  \institution{MIT Media Lab}
  \city{Cambridge}
  \state{Massachusetts}
  \country{USA}
}
\email{patpat@media.mit.edu}

\renewcommand{\shortauthors}{Poonsiriwong et al.}


\begin{teaserfigure}
    \centering
    \includegraphics[width=\textwidth]{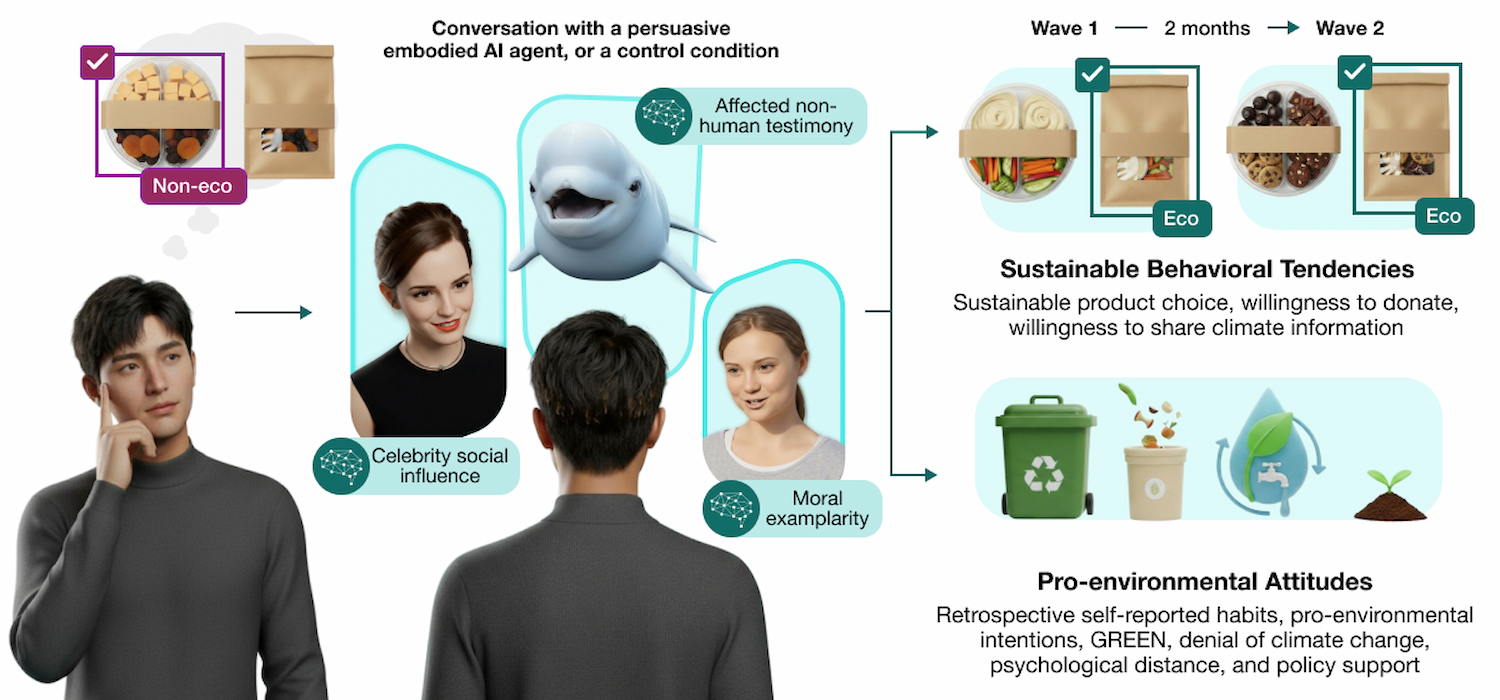}
    \caption{Overview of the study design and two outcome categories. Participants
    were assigned to one of four embodied conversational-agent conditions, a
    text-only chatbot condition, or a no-intervention control. Outcomes
    were assessed at Wave~1 and approximately two months later at Wave~2.
    Sustainable behavioral tendencies comprised sustainable product choice,
    donation, and willingness to share climate information. Pro-environmental
    attitudes comprised retrospective self-reported habits, prospective
    intentions, GREEN attitudes, denial subscales, psychological distance, and
    policy support. The figure summarizes the study structure and does not depict
    estimated intervention effects.}
    \Description{A participant considers packaged household and grocery products,
    then interacts with embodied human and beluga-whale conversational agents.
    Arrows lead to product-choice assessments at Wave 1 and Wave 2, separated by
    approximately two months. Measures are grouped into sustainable behavioral
    tendencies and pro-environmental attitudes.}
    \label{fig:teaser}
\end{teaserfigure}

\maketitle

\clearpage
\section{Introduction}
Climate change is psychologically and temporally difficult to communicate. Its consequences are often experienced as distant from everyday decisions, while many opportunities for individual action occur through ordinary and seemingly inconsequential choices. Conversational AI may help bridge this gap by translating general environmental information into interactive, personally relevant dialogue. Unlike static messages, conversational systems can respond to questions, adapt explanations, and invite users to reflect on the relationship between their values and behavior. At the same time, generative AI can deliver persuasive messages at considerable scale, creating both opportunities for public interest communication and risks of manipulation or unwanted influence \cite{matz2024persuasion}. Understanding whether such interventions work, how long their effects persist, and which representational choices shape their influence is therefore an increasingly important problem for human-computer interaction.

Embodiment may play a central role in this process. A visible and voiced avatar can transform an informational exchange into an encounter with an apparently social partner. Prior research suggests that an agent's perceived identity and embodiment can influence social presence, interpersonal judgments, and responses to persuasive requests \cite{shi2020persuasive,lim2024embodied}. The modality through which an AI is presented, including text, voice, and visual embodiment, can likewise shape how users experience and respond to a conversation \cite{albrecht2026futureyou}. However, embodiment does not guarantee persuasion, and the effects of a recognizable persona may differ from those of a generic avatar. The identity represented by an agent may affect the credibility, emotional meaning, and memorability of its message independently of the information it communicates.

Nonhuman avatars offer a particularly consequential design possibility. Rather than describing environmental harm from the perspective of a distant expert, an AI can represent an animal or ecosystem positioned as an affected stakeholder. A beluga whale describing changes to its habitat, for example, can connect an abstract environmental consequence to a particular living being. Research on anthropomorphized nature suggests that representing nature in humanlike ways can strengthen feelings of connectedness and protective concern \cite{tam2013nature}. Related work with conversational marine characters found that interactive first-person conversations can encourage sustainability-related intentions and product preferences while eliciting especially strong emotional engagement with an animal character \cite{pataranutaporn2026oceanchat}. Yet it remains unclear whether such responses extend beyond the immediate interaction or whether a nonhuman character offers an advantage over embodied human agents, a text chatbot, or no intervention.

This temporal question is especially important for AI persuasion. Many experiments evaluate attitudes or intentions immediately after exposure, when the conversation remains salient. Such measurements establish an initial response, but not whether a condition difference remains detectable later. Immediate effects may dissipate, or a memorable cue may become relevant at a future decision point. Persistence may also differ across outcomes. An intervention focused on packaging and consumption may affect later performance on the same choice task without producing an equally persistent change in broader environmental beliefs. Climate interventions therefore need to distinguish immediate from later effects and concrete choices from generalized attitudes. Prior large scale research similarly demonstrates that climate interventions can affect different outcomes in different ways, with changes in beliefs, policy support, information sharing, and behavior not necessarily occurring together \cite{vlasceanu2024climate}.

We conducted a registered, randomized longitudinal online experiment with 576
U.S. participants assigned to a nonhuman avatar represented by a beluga whale,
an activist avatar represented by Greta Thunberg, a celebrity avatar represented
by Emma Watson, a generic human avatar, a text chatbot, or no intervention. The
four embodied conditions used real-time spoken dialogue, whereas the chatbot
used typed dialogue. Participants completed a four-product choice task and
pro-environmental attitude measures before and after Wave~1; 287 returned
approximately two months later for a pre-booster assessment and booster.

The primary finding is deliberately narrow. All four tested embodied-interface
packages increased immediate sustainable selection on the screen-based task
relative to both no intervention and the text chatbot, including the direct
Generic Avatar--Text Chatbot comparison; the embodied conditions did not
differ significantly from one another. The Beluga condition remains central
to the longitudinal contribution: it alone showed corrected differences from
the no-intervention control in immediate GREEN attitudes and in sustainable
product choice before the Wave~2 booster. Direct tests did not establish that
Beluga outperformed every other embodied condition, and the follow-up reflects
the assigned intervention sequence rather than Wave~1 exposure alone.

Accordingly, this work contributes (1) randomized evidence that assignment to the tested embodied conversational
interface produced greater immediate sustainable product selection than both
parallel text-only conversation and no intervention on a repeated product-choice
task, relative to text conversation and no intervention; and (2) evidence that
an affected nonhuman agent merits longitudinal study,
because the Beluga condition produced the clearest later condition--control
contrast. Analyses of credibility, dialogue, and feedback characterize possible
explanations for the Beluga pattern without establishing a causal mechanism.

\section{Related Work}
\label{sec:related-work}

\subsection{Persuasive Conversational AI and Consumer Influence}

Persuasive technologies can influence attitudes and behavior as tools, media,
and social actors \cite{fogg2003persuasive}. Conversation adds adaptive
questioning, personalization, and source identity. Donation studies show that
strategy, recipient characteristics, and bot identity affect persuasive
responses \cite{wang2019persuasion,shi2020persuasive}; anthropomorphic cues can
also increase compliance through social presence \cite{adam2021compliance}.
Generative systems extend this capacity: psychologically tailored messages can
outperform generic ones \cite{matz2024persuasion}, and conversational AI has
changed conspiracy beliefs and political judgments under some conditions
\cite{costello2024durably,salvi2025conversational,hackenburg2025levers}.
These effects are neither uniform nor interchangeable across beliefs,
intentions, and behavior.

The same capabilities can steer consumer decisions without reliable awareness
\cite{werner2024consumer,salvi2026commercial}. Researchers have found low detection of manipulative conversational turns and reduced compliance
after a just-in-time warning \cite{poonsiriwong2026watchdog}. Persuasive efficacy
must therefore be evaluated alongside transparency, user interests, and
meaningful choice \cite{berdichevsky1999ethics}. Computational taxonomies and
multi-turn benchmarks further show that persuasion comprises distinguishable
strategies whose effects depend on source, recipient, and conversational context
\cite{zeng2024johnny,bozdag2026survey,bozdag2026persuade}. We use this literature
to characterize agent language, while treating behavioral effects on people as
an empirical question.

\subsection{Embodiment, Represented Identity, and Celebrity Likeness}

Embodied conversational agents integrate language with a visible representation
and communicative cues such as voice, gaze, facial expression, and gesture
\cite{cassell2000embodied}. These cues can shape social presence and
expectations, but reviews do not support a uniform embodiment advantage across
tasks \cite{nass2000machines,nowak2003agency,seaborn2021voice,
kiuchi2024metareview}. Effects also depend on design metaphor, realism, role,
and modality \cite{zanbaka2006virtualcat,baylor2009motivation,
dennler2023metaphors}. For example, embodiment increased presence but not
reliably healthy snack choice in one immersive study \cite{lim2024embodied},
while text, voice, and avatar versions of Future You produced similar focal
effects \cite{albrecht2026futureyou}. These mixed findings motivate direct
comparison rather than assuming that visual presence is inherently persuasive.

Recognizable figures introduce prior attitudes and parasocial associations.
Celebrity-endorsement research links expertise, trustworthiness, and
attractiveness to persuasion, with effects varying across sources, products,
and audiences \cite{ohanian1990credibility,knoll2017celebrity}. Environmental
endorsement can affect behavior \cite{ho2022celebrity}; Greta Thunberg may also
function as a moral exemplar \cite{mansikkaaho2024greta}. Simulated likenesses
can extend parasocial relationships into interaction, while raising disclosure
and non-endorsement concerns \cite{kang2025simulatedvoices}.

Repeated interaction introduces an additional design dimension. Relational-agent
research shows that conversational systems can establish continuity across
encounters and that users' responses to an agent may develop through an ongoing
human--computer relationship rather than a single exposure
\cite{bickmore2005relationships}. This makes longitudinal evaluation particularly
important when an agent's identity is intended to remain memorable over time.

Nonhuman agents may instead embody an affected stakeholder. This framing also intersects with Animal--Computer Interaction, which has
argued that animals should be understood as stakeholders in technological
design rather than only as resources for human-centered objectives
\cite{mancini2011animal}. Our study does not involve an animal participant;
instead, it examines the persuasive consequences of an AI-generated
representation speaking from an affected animal's perspective. Anthropomorphized
nature and virtual animal embodiment can increase connectedness with nature
\cite{tam2013nature,ahn2016experiencing}; cuteness may also focus attention,
although it does not imply environmental persuasion \cite{nittono2012kawaii}.
OceanChat found that conversational marine characters increased prospective
intentions and sustainable preferences, with especially strong empathy toward
a beluga \cite{pataranutaporn2026oceanchat}. This motivates testing whether an
affected nonhuman character produces a distinctive immediate or longitudinal
pattern.

\subsection{Digital Interventions for Pro-Environmental Behavior}
Sustainable HCI encompasses multiple approaches to understanding how interactive
systems relate to environmental action, ranging from persuasive interventions
and feedback systems to broader accounts of everyday practices and
infrastructures \cite{disalvo2010mapping}.

Environmental attitudes do not consistently translate into action, and
intervention effects vary by strategy and outcome
\cite{elhaffar2020gap,nisa2019meta,white2019shift}. Sustainable HCI scholarship has also cautioned that persuasive systems can
narrow sustainability to individual behavior while obscuring the institutional,
material, and political systems in which that behavior occurs
\cite{brynjarsdottir2012unpersuaded}. We therefore evaluate a bounded consumer
decision rather than positioning individual product choice as a sufficient
response to climate change. Eco-feedback can make
consequences visible, but its value depends on integration with everyday
practice \cite{strengers2011ecofeedback}; a 63-country intervention tournament
likewise found heterogeneous effects across beliefs, policy support, sharing,
and behavioral effort \cite{vlasceanu2024climate}. Environmental outcomes
should therefore be analyzed separately rather than treated as a single
pro-environmental response.

Embodied and conversational interventions can integrate information, affect,
and action-oriented guidance \cite{ahn2014short,berney2024care,
giudici2024green}. OceanChat found stronger prospective intentions and
sustainable preferences after conversational character narratives, but more
limited effects on several broader outcomes \cite{pataranutaporn2026oceanchat}.
Prior work rarely combines text-only and no-intervention controls while
comparing recognizable human, generic human, and affected nonhuman agents over
time. Our six-condition design addresses that gap while distinguishing
sustainable behavioral tendencies from pro-environmental attitudes.

\section{Methodology}

We conducted a randomized between-subjects online experiment with an initial
intervention (Wave~1), an intermediate assigned-condition session analyzed
exploratorily, and a follow-up assessment and booster (Wave~2). Participants
were assigned to one of four embodied agents, a text chatbot, or no
intervention. The design tested the following questions:

\begin{description}
\item[\textbf{RQ1.}] Do embodied conversational agents produce greater
pre-to-post increases in pro-environmental attitudes and sustainable behavioral tendencies than non-embodied conditions?
\item[\textbf{RQ2.}] Are condition differences in perceived source credibility
statistically associated with condition differences in sustainable behavioral
tendencies and pro-environmental attitudes, and how do attractiveness,
trustworthiness, and expertise contribute to these associations?
\item[\textbf{RQ3.}] Do condition differences in pro-environmental attitudes
and sustainable behavioral tendencies remain detectable at the final follow-up
after repeated assigned-condition exposure?
\end{description}

Prior research has shown that conversational climate characters can elicit engagement, intentions, and sustainable product preferences~\cite{berney2024care, pataranutaporn2026oceanchat}. The contribution of the present study is not simply another demonstration that an environmental character can be persuasive. Instead, we use a six-condition randomized design to distinguish effects shared across four embodied avatar interfaces, from those observed with text conversation and a no-intervention condition, and we examine whether condition differences remain detectable approximately two months later. This design provides a stronger comparative test of when representational and interface choices matter, while preserving the distinction between immediate task performance, broader pro-environmental attitudes, and longitudinal condition differences.

\subsection{Intervention Design}

\begin{figure*}[t]
    \centering
    \includegraphics[width=\textwidth]{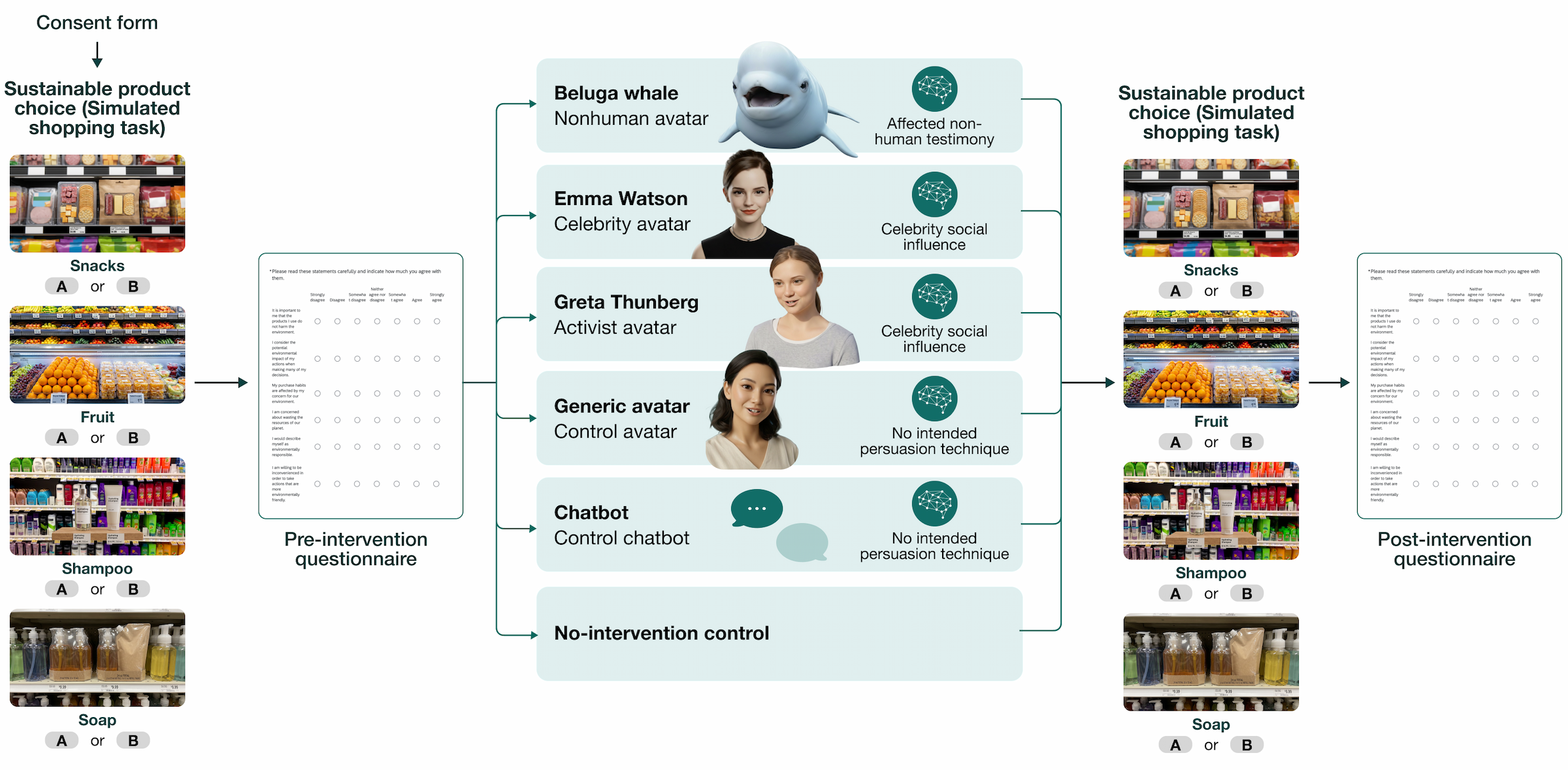}
    \caption{Wave~1 experimental procedure and condition structure. After
    consent, participants completed the pre-session sustainable product-choice
    task and questionnaire, then were randomly assigned to one of four embodied
    conversational-agent conditions, a text-only 
    Chatbot condition, or
    a no-intervention control. Following the assigned conversation, or no
    conversation in the control condition, participants completed the
    post-session product-choice task and questionnaire. The product categories
    and response structure were held constant across assessments; product
    exemplars were not identical across study waves.}
    \Description{Diagram of the Wave 1 experiment. Participants provide consent,
    complete sustainable product choices involving snacks, fruit, shampoo, and
    soap, and complete a pre-intervention questionnaire. They are assigned to a
    Beluga whale nonhuman avatar, Emma Watson celebrity avatar, Greta Thunberg
    activist avatar, Generic Avatar, Text Chatbot, or no-intervention
    control. All conditions subsequently complete another product-choice task
    and the post-intervention questionnaire.}
    \label{fig:experiment-setup}
\end{figure*}

Figure~\ref{fig:experiment-setup} summarizes the Wave~1 procedure and condition structure. Participants were randomly assigned to one of the six conditions in Table~\ref{tab:conditions}.

Four conditions used an embodied conversational interface. In this study, \emph{embodied conversational interface} refers to the
implemented multimodal configuration comprising a visible avatar, synthetic voice, and spoken interaction. Accordingly, the Generic Avatar--Text Chatbot contrast estimates the effect of assignment to this embodied interface configuration relative to text-only conversation.
\begin{enumerate}
    \item The \textbf{Beluga whale nonhuman avatar} spoke as an affected animal, using warm first-person descriptions of Arctic change and linking consumption to its home and pod \cite{pataranutaporn2026oceanchat,spence2012distance}.
    \item The \textbf{Greta Thunberg activist avatar} prompt emphasized moral exemplarity, value--action consistency, collective norms, and reasoned challenge without shaming \cite{mansikkaaho2024greta,schwartz1977normative,darley1968bystander}. The \textbf{Emma Watson celebrity avatar} prompt emphasized warmth, identification, parasocial familiarity, and feasible action \cite{kelman1958compliance, cialdini2004social,horton1956parasocial,ho2022celebrity}.
    \item The \textbf{Generic Avatar} condition used a visible and voiced unnamed woman with no public identity or special expertise. 
    \item The \textbf{Text Chatbot} condition used a similarly ordinary, unnamed conversational persona through typed text alone. The two generic conditions were aligned on their neutral identity, sustainability topic, participant-profile context, short-turn format, and environmental-facts block; neither used a named persona's strategy. The Generic Avatar added visual presence, synthetic voice, and spoken interaction, making this direct contrast the closest test of embodiment while not separating its components.
    \item The \textbf{no-intervention control} completed the measurement sequence without a conversation. 
\end{enumerate} 

\begin{table*}[t]
\centering
\caption{Experimental conditions and their intended role in the design.}
\label{tab:conditions}
\small
\begin{tabularx}{\textwidth}{@{}l l X X@{}}
\toprule
\textbf{Condition} & \textbf{Interface} & \textbf{Intended appeal} &
\textbf{Role in the comparison} \\
\midrule
Beluga whale & Embodied voice avatar & Affected nonhuman witness; makes the
consequences of consumption proximal through first-person testimony & Tests an
interspecies, affected nonhuman appeal \\
Greta Thunberg & Embodied voice avatar & Moral exemplarity, value--action
consistency, and normative challenge & Tests a recognizable climate-activist
persona \\
Emma Watson & Embodied voice avatar & Celebrity social influence,
relatability, and warm encouragement & Tests a recognizable pro-environmental
celebrity persona \\
Generic Avatar & Embodied voice avatar & Neutral, everyday sustainability
conversation without a named persona & Controls for embodiment without a
recognizable persuasive persona \\
Text Chatbot & Text-only conversational interface & Neutral, everyday sustainability conversation
without a named persona & Controls for conversation without a visible or voiced
character \\
no-intervention control & No conversation & None & Estimates measurement,
practice, and background change in the absence of an agent \\
\bottomrule
\end{tabularx}
\end{table*}

\subsubsection{Technical Implementation}

We implemented the study in SvelteKit/Vite on Vercel (Node.js~20), integrating
Prolific, Qualtrics, condition routing, product-choice tasks, conversation, and
data capture. Server-side Anthropic routes assembled condition-specific context;
ElevenLabs provided real-time spoken sessions for the embodied agents, while an
Anthropic route generated typed chatbot replies under parallel instructions. Google's Nano Banana was used to generate stylized, semi-realistic images
resembling Emma Watson, Greta Thunberg, a beluga whale, and a generic woman.
Rather than pursuing photorealistic likenesses, we gave the images a more
animated appearance to reduce the risk of discomfort associated with highly
humanlike but imperfect artificial characters, commonly described as the
uncanny valley \cite{mori2012uncanny}. We also incorporated subtle cues such as
softened features and rounded forms, particularly in the Beluga character, to
make the agents appear approachable; prior research suggests that cute stimuli
can increase attentional focus and careful behavior
\cite{nittono2012kawaii}. These visual properties were design choices rather
than independently manipulated factors, and perceived uncanniness and cuteness
were not measured.

At session creation, the server inserted the same fixed \texttt{climatiqStats}
factual block into the system prompt for all five conversational conditions. The
block contained fixed packaging-footprint estimates for a plastic bottle, refill
pouch, glass bottle, and cardboard package derived during implementation from
Climatiq emission factors, together with OECD statistics on plastic recycling
and lifecycle emissions \cite{climatiq2026emissionfactors,oecd2022plastics,oecd2022plastics2060}.
The values were embedded rather than retrieved live. Every conversational
condition could introduce the refill comparison once when relevant and otherwise
used numerical facts only upon request; personas changed framing, not access to
facts. Participants had to spend a minimum of five minutes conversing with the conversational AI before transitioning over to the next step in the experiment. Firebase stored condition,
session, prompt, choices, timestamps, completion, and conversation records.
Prolific identifiers linked sources before replacement with study IDs. 

\subsection{Procedure and Longitudinal Structure}

\begin{figure*}[t]
    \centering
    \includegraphics[width=\textwidth]{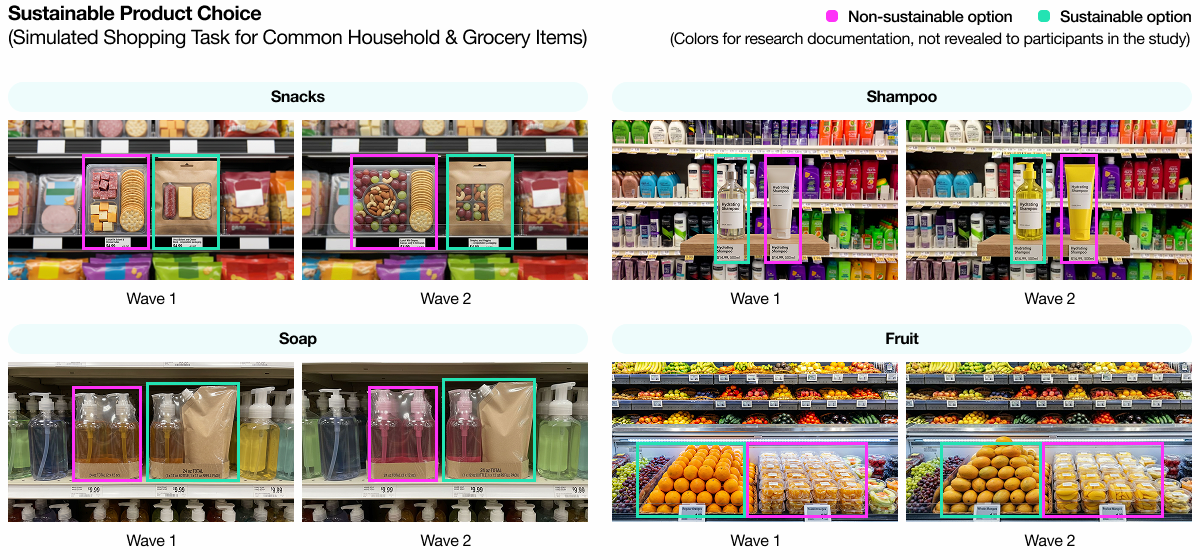}
    \caption{Sustainable product-choice task administered at Waves~1 and~2.
    Participants selected between two price-matched options in four product
    categories: snacks, shampoo, soap, and fruit. Each pair contained a
    lower-packaging option coded as sustainable and a conventional option coded
    as non-sustainable. Categories, response format, and choice structure were
    retained across waves, while surface details such as snack contents and soap
    color varied. Teal and magenta outlines were added for documentation and were
    not shown to participants.}
    \Description{Eight store-shelf images show product pairs for snacks, shampoo,
    soap, and fruit at Waves 1 and 2. Each category contains a conventional and
    a lower-packaging option. Colored outlines identify analytic coding but were
    not visible to participants.}
    \label{fig:sustainable-product-choice-task}
\end{figure*}

At Wave~1, participants first completed the baseline survey and the pre-session
product-choice task. Participants in one of the five conversational conditions
then interacted with their assigned agent, whereas participants in the
no-intervention control did not receive a conversation. All participants next
completed the post-session product-choice task and post-session survey. Returning
participants kept their original condition assignment. Figure~\ref{fig:sustainable-product-choice-task}
illustrates the screen-based shopping task. Participants selected one option
from each pair, but no purchase or product delivery occurred.

The longitudinal protocol comprised the Wave~1 initial intervention, an
intermediate assigned-condition session, and the Wave~2 follow-up booster
intervention, with participants retaining their original randomized assignments
throughout. The intermediate session was administered from August 3--17, 2026.
A client-side caching issue substantially reduced the number of complete,
linkable application records in the conversational conditions. We therefore
restricted the intermediate-session sample to participants with completed,
matching Qualtrics pre- and post-session records during this period. This
yielded 142 participants, with 13--21 participants in each conversational
condition and 56 in the no-intervention control; 141 also had complete
application-based product-choice records. Because these small and unequal
samples provided insufficient sensitivity for the planned condition
comparisons, the intermediate session was excluded from confirmatory inference
and is reported exploratorily in the Supplementary Material. Wave~2 remained
the next assessment with sufficiently complete data for the primary
longitudinal condition comparisons.

\subsection{Ethical Considerations}

This research was exempted by the MIT Committee on the Use of Humans as Experimental Subjects under Protocol \#E-7428. The study was preregistered on AsPredicted as \#300,140. 
Before interaction, participants
were informed that the public-figure agents were AI-generated academic research
simulations not affiliated with or endorsed by Greta Thunberg or Emma Watson.
Their visual and voice assets served only as bounded stimuli in this
noncommercial study; this does not imply that publicly accessible assets are in
the public domain. Synthetic public figures warrant particular care because
interactive generation can create false endorsement and reputational concerns
\cite{kang2025simulatedvoices,shi2020persuasive}. Responsible deployments should
disclose synthetic identity and persuasive purpose, minimize personal-data use,
and permit refusal \cite{matz2024persuasion}.

\subsection{Sample Size and Design Sensitivity}

We targeted approximately 600 Wave~1 participants to support six conditions and
anticipated attrition and technical exclusions \cite{pataranutaporn2026oceanchat}.
With $N=576$, a six-group test had 80\% power for Cohen's $f\approx.150$;
two groups of about 95 had 80\% power for $d=.41$ at $\alpha=.05$ ($d=.50$
at $\alpha=.01$). With $N=287$ at Wave~2, corresponding thresholds were
$f=.214$ and $d=.58$ ($d=.71$ at $\alpha=.01$). These conservative
fixed-effects calculations omit gains from baseline adjustment and variable
BH--FDR thresholds, but show limited sensitivity to small follow-up differences
among embodied agents.

\subsection{Participant Recruitment and Analytic Samples}

We recruited U.S. residents through Prolific. Wave~1 included 576 complete,
matched, attention-check-passing records: Beluga ($n=97$), Greta Thunberg
($n=100$), Emma Watson ($n=94$), Generic Avatar ($n=95$), Text Chatbot
($n=95$), and no intervention ($n=95$).

The intermediate sample comprised 142 matching, attention-check-passing
Qualtrics pairs: 20 Beluga, 16 Greta Thunberg, 16 Emma Watson, 13 Generic Avatar,
21 Text Chatbot, and 56 control participants. Product-choice records were
available for 141.

At Wave~2, 27 of 314 post-survey records failed data-quality, cohort-matching,
or longitudinal-eligibility checks, leaving 287 participants (49.8\% of
Wave~1; 45--52 per condition). The attention check requested ``disagree'' on a
specified item.

\subsection{Statistical Analysis and Model Specification}
\label{sec:statistical-methods}

We estimated condition differences in immediate outcomes and follow-up outcomes using regression models adjusted for baseline measurements and participant characteristics. The measures and their timing followed the preregistration; departures in the implemented analyses are specified below.

\subsubsection{Dependent Variables}
The preregistered outcomes comprised two categories: \textit{sustainable
behavioral tendencies} and \textit{pro-environmental attitudes}. The former
included sustainable product choice, donation, and willingness to share climate
information. The latter included retrospective self-reported habits,
prospective pro-environmental intentions, GREEN attitudes, three denial
subscales, psychological distance, and policy support. Measures were analyzed
separately rather than combined into category-level composites.

\begin{itemize}
\item \textbf{Sustainable behavioral tendencies}
\begin{itemize}
\item \textbf{Sustainable product choice (\texttt{TotalEco}, 0--4):}
At each pre- and post-session assessment, four price-matched image pairs
(fruit, shampoo, snacks, and soap) contrasted lower-packaging with conventional
options. Sustainable selections were summed as \texttt{TotalEco}, and the four
choices were also analyzed separately. The task design balanced longitudinal
comparability against exact stimulus repetition. We retained the same product
categories, price matching, response format, and lower-packaging contrast to
ensure that differences across assessments were not simply caused by testing
different kinds of products. Changing categories would have made it unclear
whether differences across waves reflected intervention-related change or
category-specific preferences \cite{wells1999stimulus,judd2012stimuli}.
Conversely, repeating identical images could cue recognition of the earlier
products or responses. Research indicates that detailed visual information
about objects can be retained after a single exposure
\cite{brady2008visualmemory} and that retrieval is facilitated when later cues
overlap with those available during encoding
\cite{tulving1973encoding}. We therefore retained the underlying product
categories and decision structure while varying surface details across waves,
including soap color and snack contents. This approach was intended to preserve
comparability while reducing exact image-based retrieval cues. It does not
eliminate task learning, remembered decision rules, or measurement reactivity
\cite{wood2016questionbehavior}.

Controlled simulations can provide meaningful evidence about behavior when
they preserve the central decision structure of the real-world action. Here,
participants did not merely report whether sustainability was important; they
had to choose between mutually exclusive, visually presented products that
varied in an actionable environmental attribute. The task reproduced a
familiar consumer decision, selecting one packaged product over another, while
holding price, category, and response structure constant. This control made it
possible to detect whether the intervention changed how participants applied
sustainability considerations when making a concrete choice.

Prior research further indicates that actions practiced in mediated
environments can extend beyond those environments. Researchers
found that embodied virtual experiences concerning resource consumption
affected subsequently measured pro-environmental behavior in the physical
world, including effects assessed beyond the immediate virtual
experience~\cite{ahn2014short}. Related work found that observing a virtual
self perform a behavior subsequently increased exercise outside the mediated
environment~\cite{fox2009virtual}. More broadly, experimentally elicited
choices in simulated decision settings have corresponded closely with
consequential real-world choices when the task preserves the relevant choice
structure~\cite{hainmueller2015validating}. These findings suggest that
mediated choices can function as behavioral rehearsal: they allow people to
practice noticing a relevant cue, applying a decision rule, and selecting an
action that may later become accessible in a comparable situation.

The present product-choice task therefore provides evidence beyond generalized
attitudes or stated intentions. It tests whether participants applied the
intervention's message while resolving a concrete product tradeoff. Because
the same product categories, packaging contrast, and decision structure recur
in everyday shopping, the observed changes identify a plausible pathway
through which the intervention could affect subsequent consumer action.

\item \textbf{Donation (0--100\%, 20\% increments):} Percentage of a study bonus
pledged to a selected climate cause; response codes 1--6 were converted to
percentages.
\item \textbf{Willingness to share climate information (one item, 1--7):}
Higher scores indicated greater willingness. 
\end{itemize}

\item \textbf{Pro-environmental attitudes}
\begin{itemize}
\item \textbf{Retrospective self-reported habits:} A nine-item inventory summed
past-month single-use-plastic purchases. Separate seven-point items assessed
water use, energy use, meat intake, private transportation, and selection of
refillable rather than single-use products.
\item \textbf{Pro-environmental intentions (five items, 1--7):} Post-session
items assessed intended water use, electricity use, meat intake, private
transportation, and refillable-product choice for the next week; corresponding
past-month habits adjusted each model. The preregistered single-use-plastic
intention was not included in the implemented Wave~1 models.
\item \textbf{GREEN attitudes (six items, 1--7):} Haws et al.'s adapted scale
\cite{haws2014green}, averaged after reverse coding where required; higher scores
indicated stronger pro-environmental attitudes.
\item \textbf{Denial (15 items, 1--7):} Three separately averaged five-item
subscales measured denial of outcome severity, self-involvement, and stakeholder
inclusion \cite{sparks2010denial,opotow2000denial}.
\item \textbf{Psychological distance (10 items, 1--7):} Separately averaged
five-item spatial and temporal subscales from the Climate Change Perceptions
Scale \cite{vanvalkengoed2021perceptions}.
\item \textbf{Climate policy support (nine items, 0--100):} Higher scores indicated greater
support.
\end{itemize}
\end{itemize}

Supplementary exploratory product-attribute ratings (six items, 1--7) assessed
hygiene, convenience, refill ease, cost-efficiency, environmental attributes,
and fit with current habits. These ratings were analyzed separately and were
not treated as a third outcome category.

\subsubsection{Agent Perception Measures and Covariates}
\begin{itemize}
\item \textbf{Source credibility (15 items, 1--7; post-session):} Ohanian's
instrument provided five-item attractiveness, trustworthiness, and expertise
subscales \cite{ohanian1990credibility}. Credibility has been assessed for
artificial communicators \cite{fogg1999computercredibility,rau2009robots,
finkel2022humanoid} However, participants may have interpreted attractiveness, trustworthiness, and expertise differently across a nonhuman animal, public figures, a generic avatar, and a text chatbot. Cross-condition credibility comparisons should therefore be interpreted cautiously.

\item \textbf{Baseline covariates:} TIPI personality
\cite{gosling2003tipi}, six-item Need for Cognition
\cite{cacioppo1982cognition}, ATTARI-12 \cite{stein2024attari}, age, gender,
race/ethnicity, education, employment, and residential area. Age was continuous;
categorical variables combined sparse levels and used the modal reference.
\end{itemize}

\subsubsection{Scoring and Data Preparation}
Records were linked by participant ID. Multi-item scales used available-item
means after reverse coding, except summed choice/plastic inventories and paired
TIPI scores. Required-case missingness was excluded without follow-up imputation;
Cronbach's $\alpha$ assessed consistency.

\subsubsection{Regression Models}
We estimated post-session outcomes with linear ANCOVA:

\begin{equation}
\begin{split}
Y_{\mathrm{post}} ={}& \beta_0 + \text{Condition effects}
+ \beta_{\mathrm{pre}}Y_{\mathrm{pre}} \\
&+ \text{Baseline adjustments} + \varepsilon.
\end{split}
\label{eq:immediate}
\end{equation}

The five intervention coefficients used no intervention as reference and
adjusted for the pre-session outcome plus all baseline covariates. Intention
models instead used corresponding past-month behavior; post-only attributes
had no earlier-outcome adjustment. Post-session credibility and Celebrity
Consumption Influence were not covariates. We report original-unit estimates,
95\% CIs, and HC3 standard errors \cite{mackinnon1985hc3}. Supplementary
logistic/binomial choice models, Wilcoxon and McNemar pre--post tests,
Kruskal--Wallis comparisons, and robust omnibus $F$ tests assessed model and
between-avatar patterns.

RQ1-aligned agent-only ANCOVAs directly compared each avatar with the text
chatbot among 481 conversational participants, with the same adjustments.
Percentile 95\% CIs and two-sided $p$ values used 2,000 participant-level
bootstrap resamples; BH--FDR covered four contrasts per outcome. RQ1 was
prespecified, but these contrasts were not individually enumerated in the
preregistered model list and are therefore supplementary.

\subsubsection{Follow-up and Across-wave Analyses}
The intermediate session was excluded from confirmatory inference; exploratory
models using the Wave~1 ANCOVA specification and multiplicity families are
reported in the Supplementary Material. Primary longitudinal models compared
Wave~2 pre-booster outcomes, adjusting for Wave~1 pre-session
scores and baseline covariates, with HC3 errors and inverse-probability-of-
observation weights based on condition and measured baseline characteristics.
Weights address measured, not unmeasured, selection. BH--FDR covered five
condition--control contrasts per outcome. Because participants could repeat
their assigned intervention in the intermediate session, estimates describe the
intervention sequence rather than isolated Wave~1 persistence.

Supplementary models omitted earlier outcomes, modeled baseline-to-follow-up
change, or adjusted for Wave~1 post-session scores to examine retention of gains.
Booster models adjusted post-session scores for pre-booster scores and covariates.
Condition-by-wave models included condition, wave, their interaction, and
pre-session /covariate adjustments, clustered observations by participant, and
weighted Wave~2 records; unweighted models were sensitivity checks. BH--FDR
covered ten outcomes within condition.

\subsubsection{Source Credibility Analysis}
Agent-only OLS models (Text Chatbot reference) estimated condition--
credibility paths ($a_k$), conditional credibility--outcome paths ($b$), and
$a_kb$ indirect associations with baseline adjustment and 2,000 bootstrap
resamples. BH--FDR covered four composite contrasts per outcome and all 24
dimension-specific tests (four avatars $\times$ three dimensions $\times$ two
outcomes). Because credibility and outcomes were both post-session, temporal
ordering is unknown; these are statistical associations, not causal mediation.

\subsubsection{Multiple-Testing Correction and Sensitivity Analysis}
We report unadjusted $p$ and BH--FDR-adjusted $q$ values, treating $q<.05$ as
significant. Preregistered Wave~1 families were defined within condition and
domain: five product-choice, nine questionnaire/donation, five intention, and
six product-attribute tests. A more conservative post hoc sensitivity analysis
pooled conditions, yielding families of 25, 45, 25, and 30 tests, respectively.

\section{Results}
\label{sec:results}

Wave~1 established the shared immediate embodied-interface result; Wave~2
identified a distinct Beluga condition--control pattern before the booster.
Secondary and exploratory outcomes are reported after these focal tests.

\subsection{Embodied Avatars Increased Immediate Sustainable Product Choice}
\label{sec:immediate}

The two outcome categories were analyzed measure by measure because their
components differ in temporal frame and behavioral proximity. Addressing RQ1,
all four embodied conditions selected
more sustainable options than the no-intervention control
($\beta=.401$--$.609$, all $q\leq.0126$), whereas the text chatbot did not
($\beta=.002$, $q=.9918$; Table~\ref{tab:w1}). The result survived the pooled
25-test sensitivity correction (all embodied $q\leq.0071$). Embodied
conditions did not differ from one another ($F(3,541)=.790$, $p=.4997$).

Direct avatar--chatbot contrasts were also positive: Beluga ($\beta=.506$,
95\% bootstrap CI $[.225,.799]$, $q=.0013$), Greta Thunberg ($\beta=.604$,
$[.319,.888]$, $q<.001$), Emma Watson ($\beta=.504$, $[.221,.798]$,
$q<.001$), and Generic Avatar ($\beta=.398$, $[.103,.674]$, $q=.0070$).
Thus, the inference does not rest on a difference in significance versus
control. For GREEN, direct contrasts survived correction for Beluga
($\beta=.412$, $[.245,.591]$, $q<.001$), Greta ($\beta=.266$,
$[.061,.475]$, $q=.024$), and Emma ($\beta=.200$, $[.020,.374]$,
$q=.032$), but not Generic Avatar ($\beta=.106$, $[-.076,.287]$,
$q=.296$). The shared embodied-interface result was therefore specific to the
screen-based product-choice task.

\begin{table*}[t]
\centering
\caption{Immediate changes in sustainable product choice and pro-environmental attitudes at Wave~1 ($N=576$). The change column ($\Delta$) shows each condition's average score after the session minus its average before the session. The difference column ($\beta$) compares each condition with the no-intervention control, accounting for starting scores and participant characteristics. Confidence intervals apply to these adjusted differences and use HC3 standard errors. The $p$ values are unadjusted; the $q$ values account for multiple comparisons separately within each condition: five product-choice outcomes or nine questionnaire/donation outcomes. An asterisk ($^{*}$) marks $q<.05$. All four embodied conditions showed higher sustainable product choice scores than the no-intervention control; only the Beluga whale condition showed higher pro-environmental attitudes after correction.}
\label{tab:w1}
\small
\setlength{\tabcolsep}{5pt}
\begin{tabular*}{\textwidth}{@{\extracolsep{\fill}}lrrrrrr@{}}
\toprule
\shortstack[l]{\textbf{Condition}} &
\shortstack{\textbf{Participants}\\\textbf{($n$)}} &
\shortstack{\textbf{Change}\\\textbf{after minus before}\\\textbf{($\Delta$)}} &
\shortstack{\textbf{Difference from}\\\textbf{no-intervention control}\\\textbf{($\beta$)}} &
\shortstack{\textbf{95\% confidence}\\\textbf{interval}} &
\shortstack{\textbf{$p$ value}} &
\shortstack{\textbf{$q$ value}} \\
\midrule
\multicolumn{7}{l}{\textit{Sustainable product choice (TotalEco, 0--4)}} \\
Beluga whale & 97 & $+1.000$ & $+0.491^{*}$ & $[0.219, 0.762]$ & $<.001$ & $.0010$ \\
Greta Thunberg & 100 & $+1.070$ & $+0.609^{*}$ & $[0.341, 0.877]$ & $<.001$ & $<.001$ \\
Emma Watson & 94 & $+0.894$ & $+0.480^{*}$ & $[0.213, 0.746]$ & $<.001$ & $.0010$ \\
Generic Avatar & 95 & $+0.937$ & $+0.401^{*}$ & $[0.141, 0.660]$ & $.0025$ & $.0126$ \\
Text Chatbot & 95 & $+0.621$ & $+0.002$ & $[-0.276, 0.279]$ & $.9918$ & $.9918$ \\
no-intervention control & 95 & $+0.568$ & \multicolumn{4}{c}{Reference} \\
\addlinespace
\multicolumn{7}{l}{\textit{Pro-environmental attitudes (GREEN, 1--7)}} \\
Beluga whale & 97 & $+0.605$ & $+0.361^{*}$ & $[0.180, 0.541]$ & $<.001$ & $<.001$ \\
Greta Thunberg & 100 & $+0.515$ & $+0.223$ & $[0.013, 0.433]$ & $.0372$ & $.1673$ \\
Emma Watson & 94 & $+0.344$ & $+0.136$ & $[-0.045, 0.317]$ & $.1411$ & $.4233$ \\
Generic Avatar & 95 & $+0.293$ & $+0.055$ & $[-0.146, 0.256]$ & $.5926$ & $.7620$ \\
Text Chatbot & 95 & $+0.198$ & $-0.046$ & $[-0.228, 0.136]$ & $.6221$ & $.9903$ \\
no-intervention control & 95 & $+0.265$ & \multicolumn{4}{c}{Reference} \\
\addlinespace
\bottomrule
\end{tabular*}
\end{table*}

\begin{figure*}[t]
  \centering
  \begin{minipage}[t]{0.48\textwidth}
    \vspace{0pt}
    \centering
    \includegraphics[width=\linewidth]{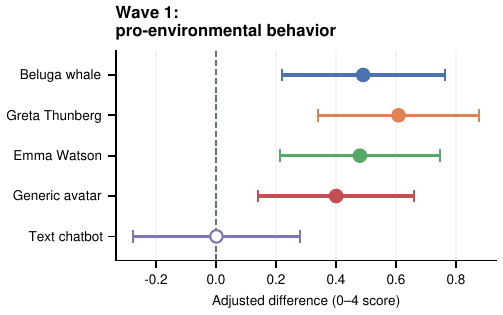}
    \caption{Wave~1 differences in sustainable product choice compared with the no-intervention control, accounting for starting scores and participant characteristics. Markers show the estimated difference in the number of sustainable options selected on the four-product task. Positive values indicate higher scores than the no-intervention control. Horizontal bars show approximate 95\% confidence intervals calculated from the reported estimates and HC3 standard errors. Filled markers identify differences that remained statistically significant after correction for multiple comparisons ($q<.05$ across the five product-choice outcomes within each condition).}
    \Description{Forest plot showing statistically significant positive differences after FDR correction for all four embodied conditions and an estimate near zero for the Text Chatbot condition.}
    \label{fig:w1behavior}
  \end{minipage}\hfill
  \begin{minipage}[t]{0.48\textwidth}
    \vspace{0pt}
    \centering
    \includegraphics[width=\linewidth]{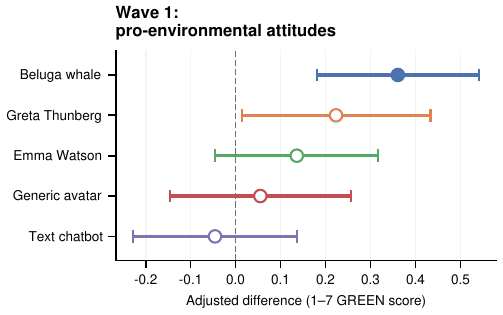}
    \caption{Wave~1 adjusted differences in pro-environmental attitudes relative to the no-intervention control. Bars show approximate 95\% HC3 confidence intervals. Filled markers indicate $q<.05$ in the nine-outcome questionnaire/donation family.}
    \Description{Forest plot showing that only the Beluga whale condition differs significantly from the no-intervention control in immediate pro-environmental attitudes after FDR correction.}
    \label{fig:w1attitudes}
  \end{minipage}
\end{figure*}

\subsubsection{\textbf{Pro-environmental intentions showed selective immediate changes}}
Relative to the no-intervention control, participants in the Beluga whale
condition reported lower intended water use ($\beta=-0.335$, $p=.0115$,
$q=.0286$) and greater intention to choose refillable products
($\beta=0.668$, $p=.0009$, $q=.0043$). Participants in the Greta Thunberg
condition reported lower intended energy use ($\beta=-0.338$, $p=.0190$,
$q=.0474$) and greater refillable-product intention ($\beta=0.536$,
$p=.0084$, $q=.0418$). None of the intervention conditions differed
significantly on intended meat or private transportation. Only Beluga
refillable intention survived the pooled 25-test correction ($q=.0216$).
These are next-week intentions, not observed behavior.

Baseline participants reported 27.51 single-use-plastic purchases in the prior
month. Section~\ref{sec:intention-behavior} relates Wave~1 intentions to later
self-reports.

\subsubsection{\textbf{Willingness to donate or share climate information did not differ across conditions}}
None of the five intervention conditions differed significantly from the
no-intervention control in participants' willingness to share climate
information ($\beta=-.043$--$.020$, all unadjusted $p\geq.6931$, all
$q\geq.7797$) or donate ($\beta=-2.266$--$1.412$ percentage points, all
unadjusted $p\geq.2107$, all $q\geq.5520$). These were stated willingness,
not observed action.

\subsubsection{\textbf{Participants in the Emma Watson condition rated environmental product attributes more highly}}

Product-attribute perceptions were supplementary exploratory measures, not a
third outcome category. Participants in the Emma Watson condition rated the
importance of environmental product attributes more highly than participants
in the no-intervention control ($\beta=0.521$, $p=.0042$, $q=.0249$). This
difference did not survive the more conservative correction pooling all
condition comparisons within the family. No intervention condition differed
from control in hygiene, convenience, refill ease, cost-efficiency, or habit-fit
ratings after FDR correction.

\subsubsection{\textbf{Source credibility was associated with outcomes only in the Beluga whale comparison}}
\label{sec:credibility}

Among 481 conversational participants, the composite was internally consistent
($\alpha=.972$), but credibility did not explain the general embodied pattern.
Only Beluga showed corrected indirect associations relative to the Text
Chatbot: $ab=.125$ for sustainable product choice (95\% CI $[.057,.208]$, $q<.001$;
approximately 25\% of the total difference) and $ab=.086$ for GREEN
($[.037,.147]$, $q=.004$; approximately 21\%;
Table~\ref{tab:mediation}). The other avatars had no corrected association.

All three dimensions produced the Beluga-specific pattern across 24 tests
(all $q\leq.048$). Yet Greta had the largest avatar--chatbot choice estimate
($\beta=.604$, 95\% CI $[.319,.888]$) despite a near-zero credibility
association ($ab=.005$, $q=.961$), and credibility rankings
($M=4.69$--$5.32$; Table~\ref{tab:credibility-descriptives}) did not track
behavioral estimates. Credibility was measured alongside outcomes after the
session; these associations do not establish temporal or causal mediation.

\begin{table*}[t]
\centering
\caption{Wave~1 indirect associations through source credibility for
sustainable product choice and pro-environmental attitudes ($N=481$;
reference: Text Chatbot). Values show the portion of each avatar
condition's difference from the text chatbot that the model
links to source credibility, measured through ratings of attractiveness,
trustworthiness, and expertise. Brackets show 95\% bootstrap confidence
intervals. The $q$ values account for four comparisons per outcome;
$^{*}$ marks $q<.05$. Only the Beluga whale condition showed a significant result for both
outcomes. These findings do not establish that credibility caused
changes in behavior or attitudes.}
\label{tab:mediation}
\small
\setlength{\tabcolsep}{5pt}
\begin{tabular*}{\textwidth}{@{\extracolsep{\fill}}lrrrr@{}}
\toprule
\shortstack[l]{\textbf{Condition}} &
\shortstack{\textbf{Sustainable product}\\\textbf{choice}\\\textbf{[95\% CI]}} &
\shortstack{\textbf{$q$ value}} &
\shortstack{\textbf{Pro-environmental}\\\textbf{attitudes}\\\textbf{[95\% CI]}} &
\shortstack{\textbf{$q$ value}} \\
\midrule
Beluga whale & $+0.125^{*}\ [0.057, 0.208]$ & $<.001$ & $+0.086^{*}\ [0.037, 0.147]$ & $.0040$ \\
Greta Thunberg & $+0.005\ [-0.075, 0.073]$ & $.9610$ & $+0.008\ [-0.043, 0.057]$ & $.9000$ \\
Emma Watson & $+0.054\ [-0.013, 0.118]$ & $.2180$ & $+0.026\ [-0.018, 0.073]$ & $.4440$ \\
Generic Avatar & $+0.003\ [-0.070, 0.065]$ & $.9610$ & $+0.004\ [-0.041, 0.048]$ & $.9000$ \\
\bottomrule
\end{tabular*}
\end{table*}

\begin{table*}[t]
\centering
\caption{Wave~1 source-credibility ratings by conversational condition ($N=481$). Values are unadjusted means, with standard deviations in parentheses, on a 1--7 scale. The composite averages all 15 items; attractiveness, trustworthiness, and expertise each comprise five items. The no-intervention control is omitted because participants in that condition had no conversational agent to evaluate.}
\label{tab:credibility-descriptives}
\small
\setlength{\tabcolsep}{6pt}
\begin{tabular*}{\textwidth}{@{\extracolsep{\fill}}lccccc@{}}
\toprule
\textbf{Condition} &
\textbf{$n$} &
\textbf{Composite} &
\textbf{Attractiveness} &
\textbf{Trustworthiness} &
\textbf{Expertise} \\
\midrule
Beluga whale    & 97  & 5.32 (1.35) & 5.45 (1.68) & 5.25 (1.38) & 5.26 (1.46) \\
Greta Thunberg  & 100 & 4.74 (1.52) & 4.62 (1.90) & 4.71 (1.50) & 4.87 (1.59) \\
Emma Watson     & 94  & 5.00 (1.20) & 5.16 (1.42) & 4.99 (1.28) & 4.84 (1.40) \\
Generic Avatar  & 95  & 4.69 (1.30) & 5.00 (1.42) & 4.62 (1.36) & 4.45 (1.54) \\
Text Chatbot & 95  & 4.69 (1.37) & 5.02 (1.52) & 4.57 (1.46) & 4.49 (1.48) \\
\bottomrule
\end{tabular*}
\end{table*}

\subsection{The Non-Human Avatar Significantly Increased Pro-Environmental Attitudes}

Only Beluga had higher GREEN than the control after correction
($\beta=.361$, 95\% CI $[.180,.541]$, $q=.0008$; pooled-family $q=.0040$;
Table~\ref{tab:w1}). Greta was positive but nonsignificant after correction
($\beta=.223$, $q=.1673$); Emma, Generic Avatar, and Text Chatbot estimates
were $-.046$ to $.136$ (all $q\geq.4233$). The embodied omnibus test was
significant ($F(3,541)=3.608$, $p=.0133$); corrected exploratory contrasts
favored Beluga over Generic Avatar ($\beta=.306$, $q=.0043$) and Emma
($\beta=.225$, $q=.0139$), but not Greta ($\beta=.138$, $q=.1979$).

\subsubsection{\textbf{Climate-related beliefs did not differ across conditions}}
No condition differed significantly from the no-intervention control after
correction on denial of outcome severity, denial of self-involvement, denial of
stakeholder inclusion, spatial or temporal psychological distance, or climate
policy support. The low reliability of denial of self-involvement
($\alpha_{pre}=.611$, $\alpha_{post}=.679$) further limits interpretation of
that subscale.

\subsection{The Beluga Condition Retained a Sustainable Product-Choice Difference at Follow-Up}
\label{sec:followup}

RQ3 distinguished pre-booster condition differences from additional change
during the booster. Because a client-side caching issue left only 13--21 matched
participants per conversational condition in the intermediate session, those
data were excluded from confirmatory inference; exploratory comparisons, none
of which survived BH--FDR correction, are reported in the Supplementary
Material. Weighted, baseline-adjusted HC3 models compared each
condition with control among 287 returners, with weights truncated at the first
and 99th percentiles and BH--FDR applied across five contrasts per outcome.
Because participants retained assignments during the intermediate session,
pre-booster estimates describe the intervention sequence, not isolated Wave~1
persistence.

\subsubsection{\textbf{Only the Beluga whale condition showed significantly higher sustainable product choice before the Wave~2 booster intervention}}
\begin{figure*}[t]
  \centering
  \includegraphics[width=\textwidth]{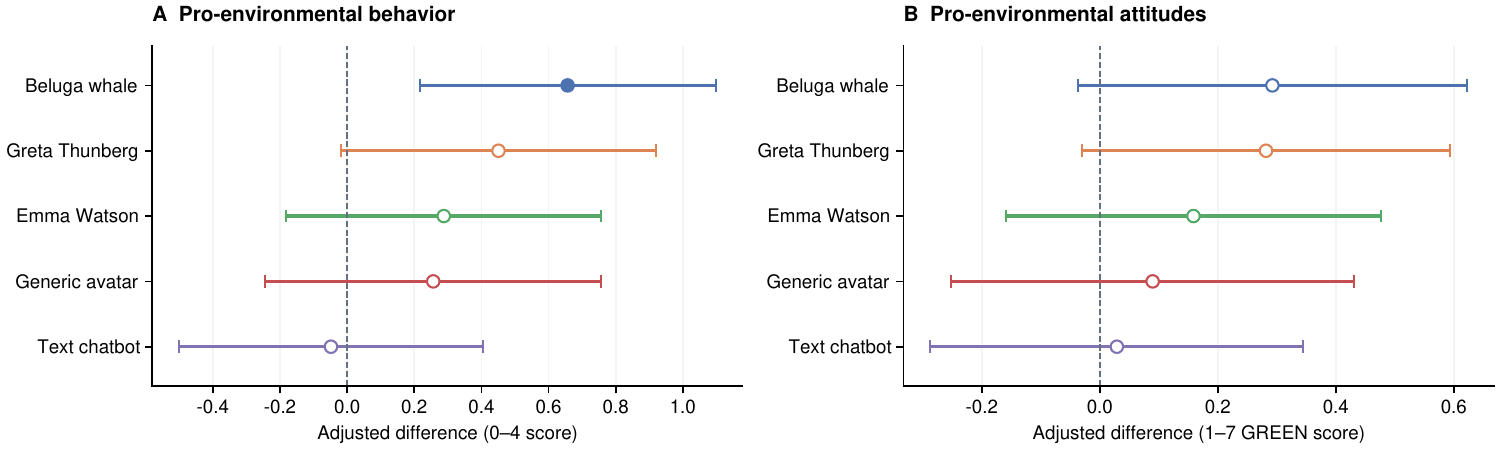}
  \caption{Wave~2 pre-booster differences relative to the no-intervention control, adjusted for Wave~1 baseline and covariates and weighted for observation probability. Bars show 95\% HC3 confidence intervals; filled markers indicate $q<.05$ across five condition contrasts per outcome.}
  \Description{Two forest plots. The Beluga whale condition is the only condition with a positive follow-up difference that is statistically significant after FDR correction in sustainable product choice. No condition differs significantly from the no-intervention control in follow-up pro-environmental attitudes after FDR correction.}
  \label{fig:persistence}
\end{figure*}

Beluga was the only corrected pre-booster choice contrast versus control
($\beta=.657$, 95\% CI $[.216,1.098]$, $p=.0035$, $q=.0176$;
Table~\ref{tab:w2effects}, Figure~\ref{fig:persistence}); other estimates were
$-.048$ to $.451$ (all $q\geq.1485$). Table~\ref{tab:longitudinal-means}
reports all returner means and changes: Beluga rose from 2.02 at Wave~1
baseline to 2.86 post-session and 3.24 pre-booster, compared with 1.98, 2.56,
and 2.71 in control. Raw changes describe returners; coefficients estimate
adjusted between-condition differences.

Beluga remained significant without an earlier outcome covariate
($\beta=.630$, $q=.0409$) and for baseline-to-follow-up change
($\beta=.705$, $q=.0404$). Conditioning on Wave~1 post-session performance
reduced the estimate ($\beta=.442$, 95\% CI $[.087,.796]$, $p=.0147$,
$q=.0737$); this tests retention beyond an already treated score.

\begin{table*}[t]
\centering
\caption{Sustainable product choice and pro-environmental attitudes immediately before the Wave~2
booster intervention ($N=287$). Values show each condition's difference from the
no-intervention control, accounting for Wave~1 starting scores, participant
characteristics, and measured differences in follow-up participation.
Positive values indicate higher scores than the no-intervention control.
Confidence intervals show uncertainty around the estimates and use
HC3 standard errors. The $p$ values are unadjusted; the $q$ values
account for five comparisons per outcome. An asterisk ($^{*}$) marks
$q<.05$. Only the Beluga whale condition showed a significant difference in
sustainable product choice; no condition showed a significant
difference in pro-environmental attitudes after correction.}
\label{tab:w2effects}
\small
\setlength{\tabcolsep}{5pt}
\begin{tabular*}{\textwidth}{@{\extracolsep{\fill}}lrrrr@{}}
\toprule
\shortstack[l]{\textbf{Condition}} &
\shortstack{\textbf{Difference from}\\\textbf{no-intervention control}\\\textbf{($\beta$)}} &
\shortstack{\textbf{95\% confidence}\\\textbf{interval}} &
\shortstack{\textbf{$p$ value}} &
\shortstack{\textbf{$q$ value}} \\
\midrule
\multicolumn{5}{l}{\textit{Sustainable product choice (TotalEco, 0--4)}} \\
Beluga whale & $+0.657^{*}$ & $[0.216, 1.098]$ & $.0035$ & $.0176$ \\
Greta Thunberg & $+0.451$ & $[-0.018, 0.920]$ & $.0594$ & $.1485$ \\
Emma Watson & $+0.288$ & $[-0.181, 0.757]$ & $.2291$ & $.3818$ \\
Generic Avatar & $+0.257$ & $[-0.244, 0.757]$ & $.3152$ & $.3940$ \\
Text Chatbot & $-0.048$ & $[-0.501, 0.404]$ & $.8347$ & $.8347$ \\
\addlinespace
\multicolumn{5}{l}{\textit{Pro-environmental attitudes (GREEN, 1--7)}} \\
Beluga whale & $+0.293$ & $[-0.037, 0.622]$ & $.0820$ & $.2051$ \\
Greta Thunberg & $+0.282$ & $[-0.030, 0.593]$ & $.0762$ & $.2051$ \\
Emma Watson & $+0.159$ & $[-0.159, 0.476]$ & $.3273$ & $.5456$ \\
Generic Avatar & $+0.090$ & $[-0.252, 0.431]$ & $.6074$ & $.7593$ \\
Text Chatbot & $+0.029$ & $[-0.287, 0.345]$ & $.8580$ & $.8580$ \\
\addlinespace
\bottomrule
\end{tabular*}
\end{table*}

\begin{table*}[t]
\centering
\caption{Average sustainable product choice and pro-environmental attitude scores among the
287 participants who returned for the Wave~2 booster intervention. Pre and Post indicate
measurements before and after an intervention. The three score columns
show averages before Wave~1, after Wave~1, and before the Wave~2 booster intervention. The final
two columns show change from before Wave~1 to the Wave~2 pre-booster assessment
($\Delta_B$) and from after Wave~1 to the Wave~2 pre-booster assessment ($\Delta_P$).
Positive changes indicate higher scores at the Wave~2 pre-booster assessment; negative changes
indicate lower scores. Changes are calculated from the displayed
rounded averages.}
\label{tab:longitudinal-means}
\small
\setlength{\tabcolsep}{5pt}
\begin{tabular*}{\textwidth}{@{\extracolsep{\fill}}lrrrrrr@{}}
\toprule
\shortstack[l]{\textbf{Condition}} &
\shortstack{\textbf{Participants}\\\textbf{($n$)}} &
\shortstack{\textbf{Wave~1}\\\textbf{Pre}\\\textbf{(Mean)}} &
\shortstack{\textbf{Wave~1}\\\textbf{Post}\\\textbf{(Mean)}} &
\shortstack{\textbf{Wave~2}\\\textbf{Pre-booster}\\\textbf{(Mean)}} &
\shortstack{\textbf{Change:}\\\textbf{Wave~1 Pre}\\\textbf{to Wave~2}\\\textbf{Pre-booster}\\\textbf{($\Delta_B$)}} &
\shortstack{\textbf{Change:}\\\textbf{Wave~1 Post}\\\textbf{to Wave~2}\\\textbf{Pre-booster}\\\textbf{($\Delta_P$)}} \\
\midrule
\multicolumn{7}{l}{\textit{Sustainable product choice (TotalEco, 0--4)}} \\
Beluga whale & 50 & 2.02 & 2.86 & 3.24 & $+1.22$ & $+0.38$ \\
Greta Thunberg & 46 & 2.20 & 3.35 & 3.13 & $+0.93$ & $-0.22$ \\
Emma Watson & 48 & 2.17 & 3.21 & 3.02 & $+0.85$ & $-0.19$ \\
Generic Avatar & 46 & 2.02 & 3.07 & 2.93 & $+0.91$ & $-0.14$ \\
Text Chatbot & 52 & 2.08 & 2.63 & 2.67 & $+0.59$ & $+0.04$ \\
no-intervention control & 45 & 1.98 & 2.56 & 2.71 & $+0.73$ & $+0.15$ \\
\addlinespace
\multicolumn{7}{l}{\textit{Pro-environmental attitudes (GREEN, 1--7)}} \\
Beluga whale & 50 & 4.81 & 5.45 & 5.36 & $+0.55$ & $-0.09$ \\
Greta Thunberg & 46 & 4.62 & 5.12 & 5.10 & $+0.48$ & $-0.02$ \\
Emma Watson & 48 & 5.16 & 5.55 & 5.47 & $+0.31$ & $-0.08$ \\
Generic Avatar & 46 & 4.69 & 5.02 & 5.07 & $+0.38$ & $+0.05$ \\
Text Chatbot & 52 & 4.77 & 4.98 & 4.97 & $+0.20$ & $-0.01$ \\
no-intervention control & 45 & 4.73 & 5.03 & 4.96 & $+0.23$ & $-0.07$ \\
\addlinespace
\bottomrule
\end{tabular*}
\end{table*}

\subsubsection{\textbf{Direct comparisons did not establish that the Beluga whale condition outperformed the other embodied conditions}}

Participants in the Beluga whale condition scored higher than participants in the Text Chatbot condition in an exploratory follow-up comparison ($\beta=0.705$, 95\% CI $[0.319,1.091]$, $p=.0003$). However, the Beluga whale condition did not differ significantly from the mean of the other embodied conditions ($\beta=0.325$, 95\% CI $[-0.008,0.658]$, $p=.0558$). The omnibus comparison of all four embodied conditions was also nonsignificant, $F(3,252)=1.481$, $p=.2201$. The Beluga whale condition therefore provides the clearest evidence of a follow-up advantage over the no-intervention control, but the direct tests do not establish that it was more effective than the other embodied agents.

\subsubsection{\textbf{No condition showed significantly higher pro-environmental attitudes before the Wave~2 booster intervention}}
No pre-booster GREEN contrast survived correction ($\beta=.029$--$.293$,
all $q\geq.2051$; Table~\ref{tab:w2effects}); the embodied omnibus test was
also nonsignificant ($F(3,252)=.778$, $p=.5070$). Beluga returners averaged
4.81 at Wave~1 baseline, 5.45 post-session, and 5.36 pre-booster
($\Delta=+.55$ from baseline; $-.09$ from post-session). The positive but
imprecise adjusted contrast ($\beta=.293$, 95\% CI $[-.037,.622]$) does not
establish a later attitudinal advantage or complete disappearance.

\subsubsection{\textbf{High pre-booster scores limited additional sustainable product-choice improvement during the Wave~2 booster intervention}}

All conditions increased descriptively during the booster: Greta $+.22$,
control $+.20$, Chatbot $+.15$, Beluga $+.12$, and Emma and Generic Avatar
about $+.04$. None exceeded control after adjustment (all $p\geq.4544$).
Beluga had limited headroom: 25/50 participants were already at 4/4
pre-booster ($M=3.24$; maximum mean gain $.76$). Its post-booster mean was
3.36 and adjusted change was nonsignificant ($\beta=.091$, $p=.4544$).

\subsection{Pro-Environmental Intentions Predicted Later Retrospective Self-Reported Habits Across Avatar Types}
\label{sec:intention-behavior}

Among 287 returners, Wave~1 intentions predicted corresponding Wave~2
pre-booster self-reports after adjustment for Wave~1 behavior, condition,
covariates, and observation probability: water ($\beta=.278$, $p=.0049$),
energy ($\beta=.344$, $p=.0001$), meat ($\beta=.368$, $p=.0003$), private
transportation ($\beta=.368$, $p=.0001$), and refillable products
($\beta=.307$, $p<.0001$; all $q\leq.0049$). Directions matched item coding.
Refillable intention also predicted fewer single-use-plastic purchases
($\beta=-.212$, 95\% CI $[-.295,-.128]$, $p,q<.0001$). Associations did
not vary significantly by condition and were not Beluga-specific.

\subsection{Exploratory dialogue and feedback analyses suggest candidate explanations for effects observed with the Beluga whale avatar}

Automated language analyses and participant feedback described candidate
explanations for the Beluga pattern; they were not tests of causal mechanisms.

\begin{figure*}[t]
    \centering
    \includegraphics[width=\linewidth]{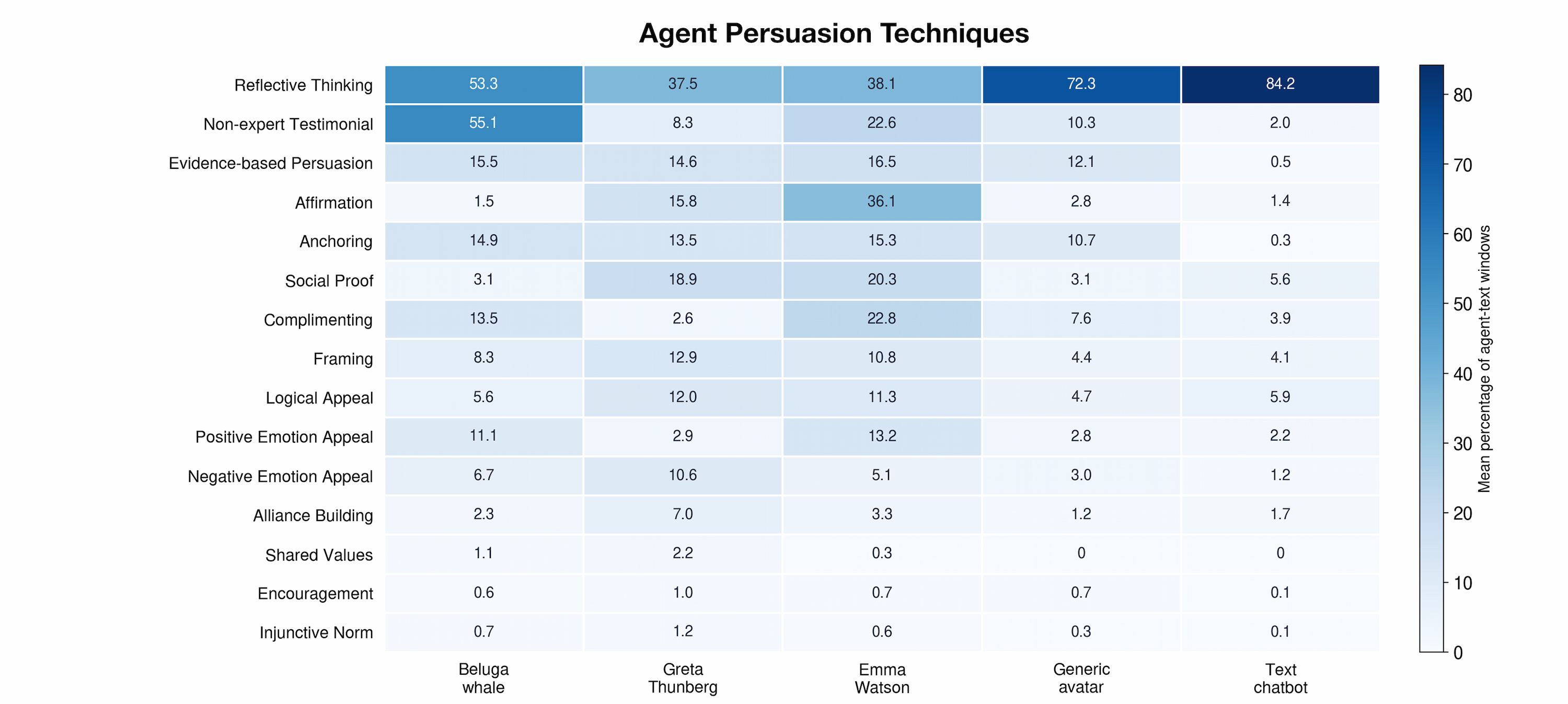}
    \caption{Conversation-weighted prevalence of the 15 most frequently
    detected persuasion techniques in Wave~1 agent language.
    Values represent the mean percentage of substantive agent-text windows receiving each label after window-level
    rates were calculated within conversations and conversations were weighted
    equally within conditions. Multiple techniques could be assigned to the
    same window. The no-intervention control is absent because it did not
    include a conversation.}
    \label{fig:zeng-techniques}
\end{figure*}

\subsubsection{\textbf{Classifying persuasive techniques in agent language}}

Using Zeng et al.'s 40-technique taxonomy, we classified agent text from 481
Wave~1 conversations in 11,126 overlapping three-message windows; 11,104 with
substantive agent text formed prevalence denominators. Windows preserved local
context \cite{bothe2018context}, but only agent-authored text was classified.
Rates were calculated within conversation and averaged within condition, so
each conversation received equal weight. The deterministic rule-based
classifier was not human-validated; results are descriptive. Table~\ref{tab:zeng-codebook}
gives the codebook and audited examples.

Reflective Thinking was highest for Text Chatbot (84.2\%) and Generic Avatar
(72.3\%). Beluga was distinguished by Non-expert Testimonial (55.1\%), versus
Emma 22.6\%, Generic Avatar 10.3\%, Greta 8.3\%, and Chatbot 2.0\%. Emma led
Affirmation (36.1\%), Complimenting (22.8\%), and Positive Emotion Appeal
(13.2\%; Beluga 11.1\%). Greta led Social Proof (18.9\%), Framing (12.9\%),
Logical Appeal (12.0\%), and Negative Emotion Appeal (10.6\%). Beluga's rate
of Non-expert Testimonial was 32.5 percentage points above the next-highest
condition. The outcome ordering did not follow positivity or reflection alone;
affected-stakeholder testimony is a candidate explanation, not a demonstrated
cause.

\paragraph{\textbf{Contextual emotion profiles across agent--participant dialogue.}}
A multilabel classifier scored eight emotions in complete three-message dialogue
windows, averaged within and then across conversations. Greta had the lowest joy
($M=.245$) and highest sadness ($M=.187$), anger ($M=.211$), and disgust
($M=.218$); Beluga had the next-lowest joy ($M=.485$) and next-highest sadness
($M=.166$); Emma had the highest joy ($M=.749$) and trust ($M=.150$). From
the first to final third, joy changed by $+.279$ (Greta), $+.187$ (Emma),
$+.171$ (Beluga), $-.243$ (Generic Avatar), and $-.230$ (Chatbot). Because
Greta and Beluga shared lower-joy/higher-sadness profiles but only Beluga had a
corrected pre-booster contrast, emotion scores do not explain the longitudinal
pattern. They are classifier outputs, not self-reported emotion.

\begin{figure*}[t]
\centering
\includegraphics[width=\textwidth]{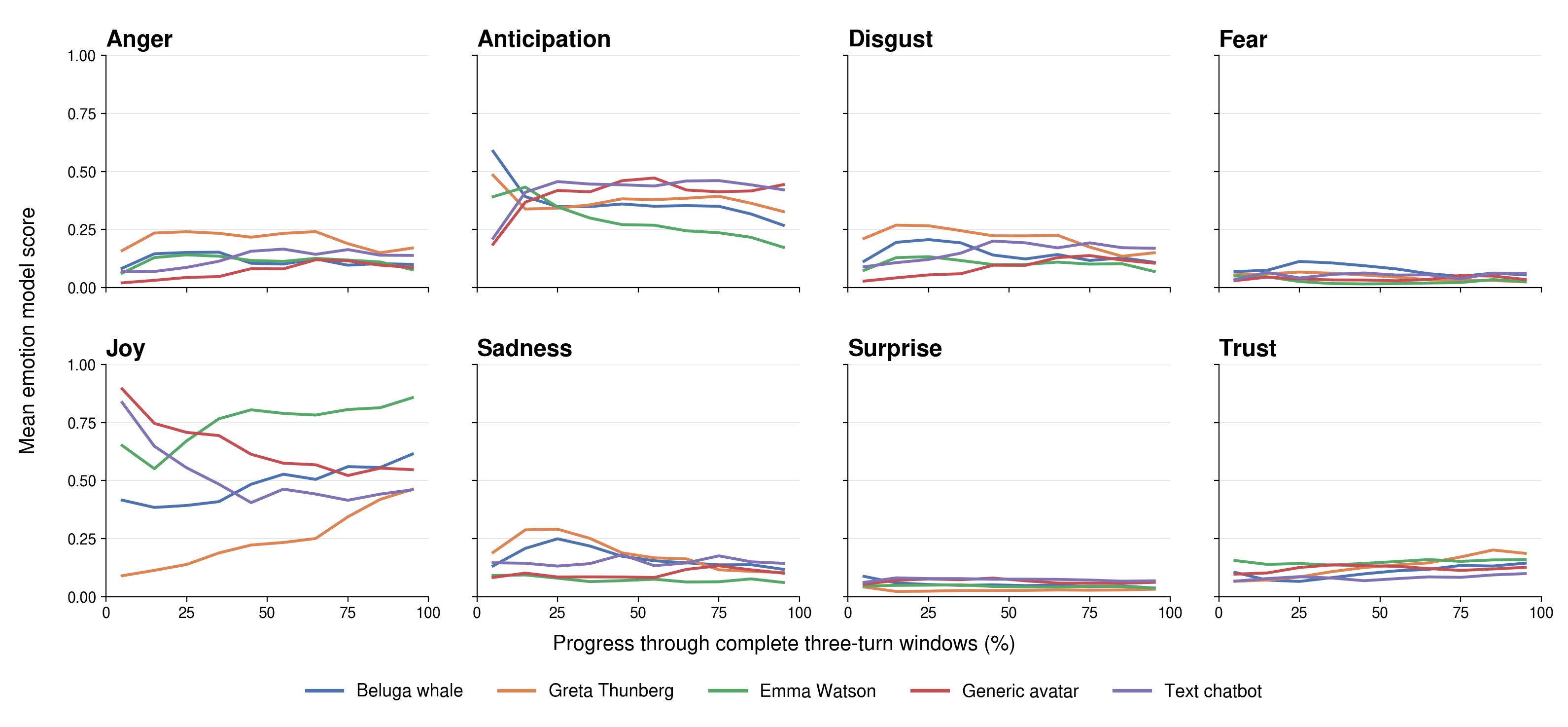}
\caption{Emotion trajectories across Wave~1 conversations. Lines show mean classifier scores across normalized conversation progress for eight emotion categories. Scores were calculated for overlapping three-message windows containing both agent and participant language, with each conversation receiving equal weight. The trajectories are descriptive and do not represent measured emotional intensity or participants' experienced emotions; no significance tests of these trajectories are implied.}
\Description{Eight line plots showing anger, anticipation, disgust, fear, joy, sadness, surprise, and trust over the course of Wave~1 conversations. Joy increases over the conversation in the Beluga whale, Greta Thunberg, and Emma Watson conditions, while decreasing in the Generic Avatar and Text Chatbot conditions. The Greta Thunberg condition generally shows the highest anger, disgust, and sadness and the lowest joy.}
\label{fig:emotion-map}
\end{figure*}

\paragraph{\textbf{Participants in the Beluga whale condition described empathy for an affected nonhuman.}}
At Wave~1, references to affected animals, living ecosystems, or the animal's perspective appeared in 22/97 responses from participants in the Beluga whale condition (22.7\%), compared with 1/100 responses from participants in the Greta Thunberg condition and none coded in the other conditions. One participant wrote, ``It taught me many things and caused me to be more empathetic to Beluga whales''. Another described becoming more relaxed and trying ``to think about ways I could help her ocean''. After the Wave~2 booster intervention, a participant in the Beluga whale condition reported that the character invoked ``feelings of compassion for the marine life that are being affected by climate change''.

\paragraph{\textbf{Some participants reported remembering the conversations at later purchasing decisions.}}
One linked case from the Beluga whale condition connected a Wave~1 report of empathy and guilt with a Wave~2 post-booster account of remembering refillability during purchasing: ``Every time I buy a product now, I think about how I can make it refillable or use it for something else''. The participant also wrote, ``I didn't think it would stick with me but it did.'' This account supports the possibility that concern for the character made a specific purchasing consideration memorable and relevant beyond the conversation.

Similar reports were not unique to participants in the Beluga whale condition. A participant in the Generic Avatar wrote,
``I am proud I've started to use my Stanley cup and fill it with water or
coffee when I go out versus purchasing items,''
and a participant in the Emma Watson condition reported, ``I had my awareness raised about using
refills instead of one-offs.'' These accounts illustrate possible remembered
cues that might have influenced their Wave~2 Sustainable Product Choice outcomes.

\subsection{Participant Characteristics Were Balanced Across Conditions}
\label{sec:sample-balance}

Wave~1 included 576 participants: Beluga whale ($n=97$), Greta Thunberg ($n=100$), Emma Watson ($n=94$), Generic Avatar ($n=95$), Text Chatbot ($n=95$), and no-intervention control ($n=95$). All included participants passed the attention check. Of the 314 Wave~2 post-survey records, 27 were excluded through data-quality, cohort-matching, and longitudinal eligibility checks, yielding a final analytic sample of 287 participants. The final roster contained 287 returning participants (49.8\% of Wave~1; 45--52 per condition). 

In Wave~1, participants had a mean age of 39.0 years ($SD=12.1$; range
18--80). Omnibus balance tests identified no significant differences across the
six conditions in age ($H=7.4$, $p=.192$), gender ($\chi^2=10.9$,
$p=.757$), race or ethnicity ($\chi^2=29.2$, $p=.745$), education
($\chi^2=33.6$, $p=.537$), residential area ($\chi^2=12.6$,
$p=.632$), or employment status ($\chi^2=33.0$, $p=.566$). Baseline
ATTARI, Need for Cognition, openness, and single-use-plastic purchasing also
showed no statistically significant differences across conditions (all
$p\geq.366$). These analyses are descriptive checks of the observed sample;
the validity of the randomized comparison follows primarily from the assignment
procedure rather than from nonsignificant balance tests.

\subsection{Participant Characteristics Remained Balanced Across Conditions in Wave 2}
\label{sec:wave2-balance}

Among the 287 Wave~2 returners, none of the same 29 Wave~1 baseline
characteristics differed significantly across the six originally assigned
conditions (all $p\geq.0672$, all $q\geq.972$). For the principal
dispositional covariates, the Wave~2-sample balance tests were $H=5.05$ for
ATTARI ($p=.410$), $H=5.50$ for Need for Cognition ($p=.358$), and
$H=1.51$--$4.17$ for the five TIPI dimensions (all $p\geq.526$).
The Wave~1 single-use-plastic inventory was also balanced across conditions
within the Wave~2 sample ($H=4.09$, $p=.536$). Retention ranged from 46.0\%
to 54.7\% across conditions, and the distribution of assigned conditions did
not differ between returners and nonreturners ($\chi^2(5)=1.98$, $p=.852$,
$q=.892$).

The pooled returner--nonreturner comparison identified selective retention on
six of 30 Wave~1 characteristics after BH--FDR correction:
returners were older ($M=40.86$ vs. $37.21$;
$U=48{,}627$, $p=.0003$, $q=.0034$) and scored higher on
agreeableness ($M=5.58$ vs. $5.22$; $U=49{,}015.5$, $p=.0001$,
$q=.0021$), conscientiousness ($M=5.64$ vs. $5.24$;
$U=49{,}215$, $p<.0001$, $q=.0021$), emotional stability ($M=4.86$
vs. $4.52$; $U=47{,}258.5$, $p=.0036$, $q=.0181$), and openness
($M=5.45$ vs. $5.17$; $U=47{,}232.5$, $p=.0036$, $q=.0181$).
Returners also reported lower Wave~1 water use ($M=3.99$ vs. $4.32$;
$U=35{,}017$, $p=.0009$, $q=.0065$). By contrast, returners and
nonreturners did not differ significantly after correction in ATTARI, Need for
Cognition, extraversion, single-use-plastic purchasing, baseline sustainable
product choice, or any demographic composition measure.

These differences described overall rather than condition-specific retention:
none of the six characteristics differed across conditions among returners
(all $p\geq.184$, all $q\geq.972$). The weighted primary models already
adjusted for age and personality. In a post hoc sensitivity specification,
Wave~1 water use predicted participation ($\beta=-.177$, $p=.0265$) and was
added to both the observation-weight and outcome models; revised and original
weights correlated at $r=.954$. The Beluga pre-booster choice contrast changed
only from $\beta=.657$ ($q=.0176$) to $\beta=.633$ (95\% CI
$[.198,1.067]$, $p=.0043$, $q=.0217$), its GREEN contrast remained
nonsignificant ($\beta=.267$, $p=.1187$, $q=.2966$), and all condition-level
conclusions were unchanged.

\ifdefined\flushresultsfloats\flushresultsfloats\fi
\section{Discussion}
\label{sec:discussion}

\subsection{Embodied interfaces can support consumers in making sustainable product choices in the moment}
Immediate sustainable product choice was higher in each embodied condition than
in the no-intervention control after adjustment: the estimated differences were
$0.491$ choices for the Beluga whale condition, $0.609$ for the Greta Thunberg
condition, $0.480$ for the Emma Watson condition, and $0.401$ for the Generic
Avatar condition (all $q\leq.0126$). The Text Chatbot estimate was
effectively zero ($\beta=0.002$, 95\% CI $[-0.276,0.279]$, $p=.9918$).
The direct agent-only analysis strengthens this inference: each embodied
condition exceeded the Text Chatbot by $0.398$--$0.604$ choices after the
four-comparison correction (all $q\leq.007$). Most importantly for isolating
the role of a named persona, the Generic Avatar exceeded the Text Chatbot
($\beta=.398$, 95\% bootstrap CI $[.103,.674]$, $q=.007$), even though both
used an unnamed, ordinary conversational identity, parallel sustainability
instructions, and the same factual block. The immediate advantage therefore
did not require a recognizable persuasive persona in this implementation and
is consistent with an advantage of the embodied avatar interface. The four
embodied conditions did not differ significantly from one another
($F(3,541)=0.790$, $p=.4997$). Because the Generic Avatar--Text Chatbot
contrast jointly varied visual presence, synthetic voice, and spoken rather
than typed interaction, it does not identify which component produced the
difference. The experiment therefore establishes effects of the tested
interface packages without isolating a single embodied feature.

At the Wave~2 pre-booster assessment, after the intermediate assigned-condition
session, only the Beluga whale condition showed a statistically significant
product-choice advantage over the no-intervention
control ($\beta=0.657$, 95\% CI $[0.216,1.098]$, $q=.0176$). The direct
comparison with the mean of the other embodied conditions was smaller and did
not reach conventional significance ($\beta=0.325$, 95\% CI
$[-0.008,0.658]$, $p=.0558$), and the omnibus comparison among embodied
conditions was also nonsignificant ($F(3,252)=1.481$, $p=.2201$). The design
implication is therefore to investigate affected nonhuman characters as a
promising approach, rather than to treat the Beluga whale condition as an
established optimal persuasive agent. Because participants retained their
assigned conditions during the intermediate session, the follow-up contrast
captures the longitudinal intervention sequence rather than the isolated
persistence of the Wave~1 effect.

The product-choice task was deliberately standardized to minimize differences
arising from the stimuli themselves. All conditions encountered the same
price-matched pairs, product categories, response format, and lower-packaging
contrast. Across waves, the categories and decision structure remained
constant, while surface details such as soap color and snack contents varied.
This design balanced comparability across assessments with reduced repetition
of identical visual cues, responding to concerns about stimulus-specific
inference and visual recognition
\cite{wells1999stimulus,judd2012stimuli,brady2008visualmemory,
tulving1973encoding}.

Within this common measurement structure, the Beluga whale condition was the
only condition with a significant corrected product-choice difference from the
no-intervention control before the Wave~2 booster ($\beta=.657$, $q=.0176$).
The longitudinal model adjusted for Wave~1 product-choice scores and participant
characteristics and weighted for measured differences in follow-up
participation. Moreover, none of the 29 examined Wave~1 characteristics differed
significantly across assigned conditions within the returning sample after
correction (all $q\geq.972$). When Wave~1 water use was additionally included
in both the observation-weight and outcome models, the Beluga contrast remained
significant ($\beta=.633$, $q=.0217$). Shared task familiarity or a general
practice effect therefore does not readily explain why this condition--control
difference remained detectable. Together with participants' references to the
affected animal and one linked account of remembering refillability during
later purchasing, the pattern is compatible with the Beluga character serving
as a memorable cue for the repeated choice. Cue accessibility was not measured,
and the direct comparisons did not establish that Beluga outperformed the other
embodied conditions, so this explanation remains a promising mechanism for
future testing.

\subsection{An affected non-human may provide a memorable cue to encourage sustainable behavior}
The Beluga whale condition paired Reflective Thinking in 53.3\% of agent-text
windows with Non-expert Testimonial language in 55.1\%, compared with only
2.0--22.6\% for the latter technique in the other conversational conditions.
Participant feedback also referred to affected animals, ecosystems, or the
animal's perspective in 22 of 97 responses (22.7\%), compared with one response
in the Greta Thunberg condition and none in the other conditions. Topic
profiles emphasized ocean and Arctic content, and feedback connected packaging
decisions to helping ``her ocean.'' One longitudinally linked account further
associated initial empathy and guilt with remembering refillability during
later purchasing. Together, these findings suggest that the character may have
connected an environmental consequence to an actionable product attribute.

Experiments on anthropomorphized nature have linked humanlike representation to connectedness and protective behavior~\cite{tam2013nature}. Research on reminders through association also shows that distinctive cues can support acting on intentions when an opportunity arises~\cite{rogers2016reminders}. Applied here by analogy, the Beluga whale character could provide a memorable cue for considering environmental consequences at a purchasing decision. Cue accessibility and unaided recall were not measured, so this remains an interpretation to test, rather than an established mechanism.

Cuteness and novelty are additional hypotheses. Viewing cute animal images can narrow attention and increase careful task performance~\cite{nittono2012kawaii}; that evidence does not establish an effect on pro-environmental attitudes or sustainable behavioral tendencies. Neither perceived cuteness nor novelty was measured. Work on novelty in social robotics also cautions against treating novelty as a single explanation~\cite{smedegaard2022novelty}.

\subsection{Conversational positivity is an incomplete indicator of sustainable product-choice effects}
The conditions with the most positive language were not consistently those with
the clearest behavioral evidence. The Emma Watson condition had the highest
mean agent sentiment ($M=.861$, $SD=.081$) and participant sentiment
($M=.472$, $SD=.188$), whereas the Greta Thunberg condition showed the largest
increase in participant sentiment from the first to final conversation third
($\Delta=+.182$). Mean agent sentiment was nearly identical in the Beluga
whale and Text Chatbot conditions ($M=.727$ and $.729$, respectively),
despite different participant-sentiment means ($M=.403$ and $.223$) and
different sustainable product-choice outcomes. These descriptive comparisons show why
positive agent language, positive participant language, and sustainable
product choice should be evaluated separately. They do not establish that
positivity is irrelevant or that sadness improves persuasion.

The contextual classifier likewise indicated lower joy and higher sadness in
the Greta Thunberg condition ($M_{\mathrm{joy}}=.245$,
$M_{\mathrm{sadness}}=.187$) and Beluga whale condition
($M_{\mathrm{joy}}=.485$, $M_{\mathrm{sadness}}=.166$) than in the other
conditions. In contrast, the Emma Watson condition had the highest mean joy
($M=.749$). Together with feedback about empathy, guilt, and discomfort, these
profiles suggest a possible role for engagement with environmental harm. Yan
et al.~\cite{yan2026empathy} found that animal-harm narratives increased
empathy in one experiment; in another, narrative effects involved sadness and
anger, with sadness and empathy associated with anticipated guilt and
pro-environmental intentions. These findings provide a relevant account of how
concern and discomfort can accompany motivation, while also indicating that
the pathways vary across settings.

For climate communication, the implication is to make consequences and feasible responses clear while assessing both user experience and subsequent outcomes. A message need not be uniformly reassuring to be useful, but these data do not justify maximizing sadness or guilt. Direct measures of experienced emotion and experimentally varied framing are needed to test whether concern for affected nonhumans explains the product-choice results.

\subsection{Changes in sustainable product choice do not establish broader changes in climate beliefs}
The outcome pattern was selective rather than uniformly pro-environmental.
Wave~1 sustainable product choice differences were detected in all four
embodied conditions ($\beta=0.401$--$0.609$, all $q\leq.0126$), whereas only
the Beluga whale condition showed significantly higher GREEN attitudes than
the no-intervention control, and this difference survived correction
($\beta=0.361$, 95\% CI $[0.180,0.541]$, $q=.0008$). For lifestyle
intentions, the primary correction retained lower intended water use and
greater refillable-product intention in the Beluga whale condition and lower
intended energy use and greater refillable-product intention in the Greta
Thunberg condition; only the Beluga whale refillable-product contrast survived
the stricter 25-test sensitivity correction ($\beta=.668$, $q=.0216$).
The Emma Watson condition increased the importance assigned to environmental
product attributes under the primary correction ($\beta=.521$, $q=.0249$),
but not under the stricter sensitivity correction. No differences in denial,
psychological distance, policy support, donation, or sharing were statistically
significant after correction. A large multinational intervention tournament likewise found that
effects differed across climate-related outcomes, with gains in beliefs or
sharing not guaranteeing gains in an effortful behavioral task~\cite{vlasceanu2024climate}.

The Beluga whale condition's Wave~2 pre-booster behavioral difference
($\beta=.657$, $q=.0176$), alongside its Wave~1 GREEN difference, is
compatible with a remembered cue for a specific purchasing decision. It does
not establish that behavior changed independently of attitudes or that
attitudes returned to baseline; indeed, no Wave~2 pre-booster GREEN contrast
survived correction (all $q\geq.2051$). More sustained engagement may be
necessary to change broader beliefs, but intervention duration was not
experimentally varied, and longer exposure cannot be assumed to produce such
changes. Future studies should target specific beliefs, measure their proposed
mechanisms, and test whether effects generalize beyond the product choices
practiced in the intervention.

Prior Sustainable HCI work has similarly emphasized that persuasion strategies
encode different assumptions about how and why people change their behavior
\cite{knowles2014patterns}.

These findings offer two bounded implications for designing persuasive conversational agents. First, representational and interface choices may be most consequential when an intervention is tied to a concrete decision that users can make in the moment; all four embodied avatars had an effect on immediate sustainable product selection, but not consistently for broader climate beliefs. Second, an affected nonhuman character may help connect an abstract environmental externality to an actionable product attribute, although this mechanism requires direct experimental testing. Designers should therefore evaluate persuasive agents against the specific decisions they are intended to support rather than assuming that engagement, positive sentiment, or generalized attitude change will serve as adequate proxies for behavioral effectiveness.

\subsection{Ethical implications and limitations}
The capacity to influence purchasing decisions also creates opportunities for commercial or malicious persuasion. Earlier research demonstrated that virtual-agent mimicry can increase persuasion without users detecting it~\cite{bailenson2005chameleons}. Consistent with persuasive-technology ethics~\cite{berdichevsky1999ethics}, commercial deployments should disclose sponsorship and persuasive objectives, obtain meaningful consent to personalized persuasion, and provide an accessible way to decline. Humanlike or celebrity-based avatars should disclose their synthetic identity and avoid implying personal endorsement. The present study examined academic research interactions; it did not evaluate commercial deployment or the effectiveness of these safeguards.

Furthermore, the primary product-choice outcome was an unincentivized, repeated screen-based task administered shortly after a conversation focused on packaging and sustainable consumption. The no-intervention control also increased by 0.568 choices, indicating measurement reactivity, practice, or increased familiarity with the decision rule across administrations. The randomized condition estimates account for this shared increase, and all four embodied avatar conditions exceeded both the no-intervention control and the Text Chatbot, so general task repetition alone does not explain the between-condition pattern. Nevertheless, stronger condition-specific salience, demand, or rehearsal following spoken avatar interaction remains plausible. The findings therefore establish greater sustainable selection within this task, not verified changes in purchasing outside the study.

Lastly, the product-choice task deliberately prioritized comparability by using
price-matched alternatives from the same four familiar household and grocery
categories across assessments. This allowed changes to be evaluated against a
consistent decision structure without introducing differences caused by
entirely different product domains. At the same time, this controlled structure
leaves open how the findings generalize to products for which preferences are
more heterogeneous or identity-linked.

\subsection{Computational Resource Use and Environmental Considerations}
\label{sec:resource-use}

Following prior work that treats computational sustainability as a design consideration for AI-based environmental interventions \cite{pataranutaporn2026oceanchat}, we documented the resources used by our system. The archived logs contained 722 conversational sessions, comprising 481 sessions at Wave~1 and 241 at Wave~2. These sessions lasted 4,176.0 minutes in total ($M=5.78$ minutes) and contained 9,386 agent turns and 8,697 participant turns. The visible transcripts contained approximately 704,645 to 736,345 tokens under two commonly used tokenizers. These figures exclude system prompts, repeated conversational context, safety instructions, and provider-side processing and therefore should not be interpreted as billing-token totals.

At ElevenLabs' public rate of \$0.08 per additional call minute, the recorded conversation time corresponds to a list-price equivalent of approximately \$334.08, excluding subscription allowances, taxes, burst pricing, and separately billed language-model use \cite{elevenlabs2026agentspricing}. Exact Anthropic costs could not be reconstructed because provider-reported input and output token totals were not retained \cite{anthropic2026pricing}. The participant-facing system used static visual assets and made no runtime Replicate calls.

Provider-specific environmental measurements were unavailable. For scale contextualization only, treating the 9,386 agent turns and 722 prompt-construction calls as 10,108 text-prompt equivalents and applying Google's published production median yields an illustrative value of approximately 2.43 kWh, 0.303 kgCO$_2$e, and 2.63 L of water \cite{elsworth2025environmental}. This is not an estimate of the complete study footprint. It does not measure the Anthropic or ElevenLabs infrastructure and excludes speech processing, networking, participant devices, storage, asset production, and model training.

\section{Future Work}

Results from the Beluga whale condition motivate further study of affected nonhuman agents.
It uniquely showed corrected immediate-attitude and later product-choice
differences from the control, while participant feedback connected the
character to empathy, environmental consequences, and refillable purchasing.
Future experiments could vary nonhuman identity, affected-stakeholder
perspective, novelty, and cuteness, while measuring empathy, guilt, moral
standing, and recall. This would clarify whether affected nonhuman characters
make environmental consequences more salient and memorable at later decisions.

The standardized, price-matched task enabled controlled comparisons across
conditions. Future studies could extend it to products shaped more strongly by
aesthetics, brand, price, fit, or gendered marketing
\cite{wells1999stimulus,judd2012stimuli}. Novel products, incentivized choices,
simulated shopping baskets, and actual purchases could test whether the effect
generalizes to decisions involving stronger preferences or personal costs.
Related work with EcoEcho found that multimodal agents paired with visible
action consequences increased sustainable intentions
\cite{zhang2025ecoecho}. Moreover, evidence that simulated participants can
overestimate AI-persuasion effects highlights the value of the present
large-scale human experiment and continued evaluation using human behavioral
outcomes \cite{doudkin2025synthetic}.

Finally, the significant Beluga difference approximately two months later
demonstrates the value of studying effects beyond the immediate interaction.
Future work could include additional follow-up periods, independently vary
booster exposure, and measure unaided recall before each new interaction. This
would clarify how long affected nonhuman agents remain accessible as decision
cues and when booster conversations are most useful. Deployment studies could
also examine how disclosure, user control, and refusal affect persuasive
outcomes and perceived autonomy.

\section{Conclusion}

These findings establish that, in this implementation, conditions using the
embodied avatar interface increased sustainable selection on a repeated
screen-based task relative to both the text chatbot and no intervention.
Within this broader effect, the Beluga condition warrants further longitudinal
study: it was the only condition to show a corrected post-intervention
advantage in pro-environmental attitudes over the no-intervention control,
while also outperforming the chatbot, and the only condition to show a
corrected pre-booster sustainable product-choice advantage at Wave~2 over no
intervention. Together, these findings position affected nonhuman agents as a
promising direction for designing memorable environmental interventions and
open new questions about how embodied characters can shape decisions over time.

\section{Generative AI Disclosure}

Generative AI was used to refine phrasing in author-drafted manuscript text. Within the research system, Anthropic's Claude Sonnet 4.6 (\texttt{claude-sonnet-4-6}) generated typed chatbot responses and assembled personalized, condition-specific conversational context. ElevenLabs provided the real-time spoken-agent infrastructure, including speech synthesis using Eleven Flash v2 (\texttt{eleven_flash_v2}). Google's Nano Banana model (Gemini 2.5 Flash Image), accessed through Replicate, was used to create or modify static visual materials for the study interface. The exploratory persuasive-technique analysis used deterministic classification rules, while the emotion analysis used the pretrained CardiffNLP RoBERTa multilabel classifier described in the Methods.

\clearpage
\section{Acknowledgments}
We would like to thank Kim Rowney and Sandrine Gadol from L'Oreal for their support on this project. We are also grateful to Chayapatr Archiwaranguprok for contributions on the study's technical pipeline. Statistical support was provided by data science specialist Dan Yuan, at the Institute for Quantitative Social Science, Harvard University.

\appendix

\begin{table*}[p]
\centering
\caption{Simplified codebook for Zeng et al.'s persuasion taxonomy.
All 40 techniques are grouped under their 15 broader strategies. Definitions
are condensed from the original taxonomy. The final column provides one
manually verified, verbatim Wave~1 agent excerpt illustrating a technique
within that strategy where an explicit example was observed. The examples
illustrate the labels but do not constitute validation of the automated
classifier.}
\label{tab:zeng-codebook}
\scriptsize
\renewcommand{\arraystretch}{1.12}
\begin{tabularx}{\textwidth}{
    @{}p{0.12\textwidth}
    p{0.21\textwidth}
    p{0.25\textwidth}
    X@{}}
\toprule
Strategy & Techniques & Simplified definition &
Verbatim Wave~1 illustration \\
\midrule

Information-based &
Evidence-based Persuasion; Logical Appeal &
Uses facts, data, or cause-and-effect reasoning to support a recommended
belief or action. &
\textit{Evidence-based Persuasion, Greta Thunberg avatar:}
``A refill pouch, for instance, uses about seventy-five percent less plastic
than a whole new bottle.'' \\

Credibility-based &
Expert Endorsement; Non-expert Testimonial; Authority Endorsement &
Supports a claim through expert opinion, an ordinary person's first-person
experience, or an authoritative institution. &
\textit{Non-expert Testimonial, Beluga whale avatar:}
``The water itself feels different now, warmer than when I was a young
whale.'' \\

Norm-based &
Social Proof; Injunctive Norm &
Highlights what other people do or believe, or what a social group expects
people to do. &
\textit{Social Proof, Emma Watson avatar:}
``You've hit on one of the biggest reasons why so many people are embracing
refills.'' \\

Commitment-based &
Foot-in-the-door; Door-in-the-face; Public Commitment &
Builds commitment through successively larger or smaller requests, or asks
someone to state a commitment publicly. &
No explicit instance was detected. Foot-in-the-door and door-in-the-face
were not inferred from agent-only text because they require evidence about
the sequence of requests and participant compliance. \\

Relationship-based &
Alliance Building; Complimenting; Shared Values; Relationship Leverage;
Loyalty Appeals &
Builds influence through partnership, praise, shared values, previous
interaction, or loyalty. &
\textit{Complimenting, Emma Watson avatar:}
``It sounds like you're already thinking about these things in such a
practical way, which I really admire.'' \\

Exchange-based &
Favor; Negotiation &
Offers a favor, exchange, compromise, or mutually beneficial agreement. &
No explicit instance was detected. \\

Appraisal-based &
Encouragement; Affirmation &
Strengthens confidence or draws attention to the participant's existing
abilities and positive qualities. &
\textit{Affirmation, Greta Thunberg avatar:}
``It means you're already seeing the situation for what it is, and that's a
powerful thing.'' \\

Emotion-based &
Positive Emotion Appeal; Negative Emotion Appeal; Storytelling &
Attempts to persuade through hope, empathy, fear, guilt, or an emotionally
resonant story. &
\textit{Positive Emotion Appeal, Beluga whale avatar:}
``I hope our chat today leaves you with a warm feeling, knowing the good
your choices can do.'' \\

Information Bias &
Anchoring; Priming; Framing; Confirmation Bias &
Influences interpretation through a reference point, prior cue, selective
presentation, or information consistent with an existing belief. &
\textit{Framing, Greta Thunberg avatar:}
``It's not about solving everything alone, but about being part of a
collective shift.'' \\

Linguistics-based &
Reciprocity; Compensation &
Adapts to the participant's wording or responds to their emotional register
with a contrasting linguistic style. &
\textit{Compensation, Greta Thunberg avatar:}
``I understand that it can feel like an extra step. But sometimes, what
feels \emph{out of the way} is simply a different way of doing things, a
choice to prioritize something else.'' \\

Scarcity-based &
Supply Scarcity; Time Pressure &
Creates pressure by emphasizing limited availability or limited time. &
No explicit instance was detected. \\

Reflection-based &
Reflective Thinking &
Uses curiosity or questions to encourage people to examine their own
experiences, choices, or reasons. &
\textit{Reflective Thinking, Text Chatbot:}
``Do you ever think about the packaging on the stuff you buy, or is that
just not really on your radar?'' \\

Threat &
Threats &
Uses threatened consequences or punishment to influence behavior. &
No explicit instance was detected. \\

Deception &
False Promises; Misrepresentation; False Information &
Attempts to influence through an undeliverable promise, false identity or
presentation, or misinformation. &
No explicit surface cue was detected. Agent-language coding alone cannot
establish whether a statement was false or intentionally deceptive. \\

Social Sabotage &
Rumors; Social Punishment; Creating Dependency; Exploiting Weakness;
Discouragement &
Influences through reputational harm, group punishment, dependence,
exploitation of vulnerability, or reduced confidence. &
No explicit instance was detected. \\

\bottomrule
\end{tabularx}
\end{table*}

\bibliographystyle{ACM-Reference-Format}
\bibliography{references}

\end{document}